\documentclass[11pt]{article}

\usepackage[margin=1in]{geometry}
\usepackage{amsmath,amsfonts,amssymb,amsthm,mathtools,nicefrac,mymath}

\usepackage{graphicx,xcolor}
\usepackage{caption}
\usepackage{subfig}
\usepackage{url}

\usepackage{tikz}
\usepackage{pgfplots}
\usepackage{circuitikz}
\usetikzlibrary{calc}
\usetikzlibrary{positioning}
\usetikzlibrary{shapes.multipart}
\usetikzlibrary{pgfplots.groupplots}

\usepackage{tabularx,multirow,multicol,booktabs}
\newcolumntype{L}{>{\raggedright\arraybackslash}X}
\newcolumntype{C}{>{\centering\arraybackslash}X}
\newcolumntype{R}{>{\raggedleft\arraybackslash}X}

\usepackage[braket]{qcircuit}
\usepackage{environ}
\newcommand{\myqctmp}[2][0.25]{\Qcircuit @C=#2em @R=#1em @!R}
\NewEnviron{myqcircuit}[1][0.25]{\vcenter{\myqctmp[#1]{0.5} {\BODY}}}
\NewEnviron{myqcircuit*}[2]{\vcenter{\myqctmp[#1]{#2} {\BODY}}}
\newcommand{\controlsq}{*!<0em,.025em>-=-<0em>{\square}} 
\newcommand{\ctrlsq}[1]{\controlsq \qwx[#1] \qw}         

\usepackage[colorlinks,urlcolor=blue,linkcolor=blue,citecolor=blue]{hyperref}
\usepackage[capitalize,nameinlink]{cleveref}

\newcommand{\qcrank}{\textbf{QCrank}}
\newcommand{\qbart}{\textbf{QBArt}}
\newcommand{\qcrankabs}{\qcrank}
\newcommand{\qbartabs}{\qbart}

\newcommand{\myvec}[1]{\vec #1}                 
\newcommand{\dpt}{x}						    
\newcommand{\data}{\myvec{\dpt}}			    
\newcommand{\apt}{\alpha}					    
\newcommand{\angles}{\myvec{\apt}}			    
\newcommand{\datameas}{\data^{~\text{meas}}}	
\newcommand{\aptmeas}{\apt^{\text{meas}}}

\newcommand{\cnot}{CX}

\newcommand{\ceil}[1]{\lceil {#1} \rceil}

\newcommand{\UCR}[1][]{
    \ifthenelse{ \equal {#1} {} }
        {\text{UCR}}
        {\text{UCR}^{(#1)}}
}

\newcommand{\UCRy}[1][]{\UCR[#1]_{y}}
\newcommand{\pUCRy}[1][]{\text{p}\UCRy[#1]}
\newcommand{\psifrqi}[1][]{
    \ifthenelse{ \equal {#1} {} }
    {\ket{\psi_{\text{FRQI}}}}
    {\ket{\psi_{\text{FRQI}}(#1)}}
}
\newcommand{\psiqcrank}[1][]{
    \ifthenelse{ \equal {#1} {} }
    {\ket{\psi_{\text{qcrank}}}}
    {\ket{\psi_{\text{qcrank}}(#1)}}
}

\title{\bf Quantum-parallel vectorized data encodings and computations on trapped-ions and transmons QPUs}
\author{%
Jan Balewski\textsuperscript{1},
Mercy G. Amankwah\textsuperscript{1,2},
Roel Van Beeumen\textsuperscript{3},\\
E. Wes Bethel\textsuperscript{4},
Talita Perciano\textsuperscript{5},
Daan Camps\textsuperscript{1}
}
\date{\scriptsize%
\textsuperscript{1}National Energy Research Scientific Computing Center, Lawrence Berkeley National Laboratory, Berkeley, CA, USA\\
\textsuperscript{2}Department of Mathematics, Case Western Reserve University, Cleveland, OH, USA\\
\textsuperscript{3}Applied Mathematics and Computational Research Division, Lawrence Berkeley National Laboratory, Berkeley, CA, USA\\
\textsuperscript{4}Computer Science Department, San Francisco State University, San Francisco, CA, USA\\
\textsuperscript{5}Scientific Data Division, Lawrence Berkeley National Laboratory, Berkeley, CA, USA\\
}

\begin{document}

\maketitle

\begin{abstract}
Compact quantum data representations are essential to the emerging field of quantum algorithms for data analysis. 
We introduce two new data encoding schemes, \qcrankabs\ and \qbartabs, which have a high degree of quantum parallelism through uniformly controlled rotation gates. 
\qcrankabs\ encodes a sequence of real-valued data as rotations of the data qubits, allowing for high storage density.
\qbartabs\ directly embeds a binary representation of the data in the computational basis, requiring fewer quantum measurements and lending itself to well-understood arithmetic operations on binary data. 
We present several applications of the proposed
encodings for different types of data. 
We demonstrate quantum algorithms for DNA pattern matching, Hamming weight calculation,  complex value conjugation, and retrieving an $O(400)$ bits image, all executed on the Quantinuum QPU. 
Finally, we use various cloud-accessible QPUs, including IBMQ and IonQ, to perform additional benchmarking experiments.
\end{abstract}

\section{Introduction}
\label{sec:intro}

Quantum computing is believed to open doorways to novel methods and algorithms that can outperform their classical counterparts~\cite{NC2010}. 
Among the most prominent examples of quantum algorithms are Shor's prime factoring algorithm~\cite{Shor1997} and Grover's unstructured search algorithm~\cite{Grover1996}.
In addition, recent results show that quantum computers have great potential to solve problems in machine learning~\cite{Ciliberto2018,Abbas2021,huang2021power,liu2021rigorous}.
Similarly, there has been a considerable amount of work in quantum image processing~\cite{yan2020quantum,PhysRevX.7.031041,Wang2022,Jiang2019,Li2020}.

A crucial problem when designing and implementing quantum algorithms that process classical data is the \emph{data encoding problem}~\cite{lloyd2020}, which is related to how data is encoded in the quantum state of a qubit register.
In the encoding process, there is a trade-off between efficient use of the Hilbert space and the computational complexity of the algorithms leveraging the quantum representation.

The main contributions of this paper are related to this data encoding problem and how to use efficient encodings to develop quantum data analysis algorithms.
We introduce an extension of the uniformly controlled rotation~\cite{Mottonen2004} that enables concurrent execution of elementary quantum gates on the address and data qubits.
Based on this new idea, we propose two new data encodings: \qcrank\ for continuous data, and \qbart\ for discrete data in binary representation.
A second set of contributions relates to a series of experiments of data encoding and analysis demonstrated using real quantum processors at an unprecedented scale.
These experiments are applied to different types of data including images, DNA sequences, and time-series.

The three most well-known 
types of data
encoding are basis encoding, amplitude encoding, and angle encoding~\cite{Schuld2021-sp}.
Assume that the input data is an $N = 2^n$-dimensional vector $\data = \left[\dpt_0, \ldots,  \dpt_{N-1} \right]$.
\textit{Basis encoding} is mainly used when discrete data must be arithmetically manipulated in a quantum algorithm.
In this case, $\data$ is a binary string obtained from the original classical data.
For example, if the classical data is the vector $[0,1,2,3]$, then $\data = [00,01,10,11]$.
This binary string is encoded in 
the computational basis states of a qubit system, i.e., $\lvert \data \rangle = \lvert 00011011 \rangle$. 
In the case of \textit{amplitude encoding}, a (normalized) real- or complex-valued data vector $\data$ is directly encoded in a $2^n$-dimensional Hilbert space through the amplitudes of the state $\sum_i \dpt_i \ket{i}$.
Finally, in an \textit{angle encoding}, each $x_i$ in $\data$ is embedded through single-qubit rotations, for example, as $\bigotimes_i \left( \cos(\dpt_i/2) \ket{0} + \sin(\dpt_i/2) \ket{1} \right)$ in case of a Pauli-$Y$ rotation.
Other types of data encoding can be found in the literature, including quantum associative memory, qsample, and quantum random access memory~\cite{Ventura1998-ju,Schuld_undated-cc,ambainis2008}.

This work is closely related to data encodings predominantly used in quantum image processing, usually referred to as quantum image representations (QIR).
A variety of QIR methods have been developed~\cite{Yan2016}. 
The (improved) flexible representation of quantum images ([I]FRQI)~\cite{Le2011AOperations,Le2011b,Khan2019}, the (improved) novel enhanced quantum representation ([I]NEQR)~\cite{Zhang2013a,Jiang2015a}, the multi-channel representation of quantum images (MCRQI/MCQI)~\cite{Sun2011,Sun2013AnComputers}, and the (improved) novel quantum representation of color digital images ([I]NCQI)~\cite{Sang2016,Su2021-vf} are among the most powerful existing QIR methods.
In our previous work, we proposed an overarching encoding framework called QPIXL~\cite{Amankwah2022QuantumImages} that unifies all QIRs mentioned above.
In the QPIXL framework, every QIR can be written as
\begin{equation}
\ket{\psi(\data)} = \sum_i \ket{i} \otimes \ket{c_i},
\label{eq:enc}
\end{equation}
where $\ket{c_i}$
is an encoding of
the pixel colors in the qubit state and $\ket{i}$ an encoding of the
pixel positions in the qubit state~\cite{Amankwah2022QuantumImages}.
All QIRs that are commonly considered in the literature use a straightforward basis encoding for the pixel position information. 
However, the color mapping varies for different QIRs. 
For example, NEQR employs a basis encoding for the pixel color information, FRQI uses an angle encoding in a single qubit.
In contrast, IFRQI and MCRQI/MCQI use angle encodings over multiple qubits.

The second contribution of QPIXL is an asymptotically optimal quantum circuit implementation to prepare QIRs based on \emph{uniformly controlled rotation} ($\UCR$) gates~\cite{Mottonen2004}.
A $\UCR$ is a multi-parameter, multi-qubit gate acting on $n_a$ control or \emph{address qubits}
and 1 target or \emph{data qubit}. 
A UCR gate performs a single-qubit rotation of the data qubit around a fixed axis on the Bloch sphere.
Here, the rotation angle 
depends conditionally on the computational basis state of the address qubits. 
As such, it is parametrized by $2^{n_a}$ rotation angles as there are $2^{n_a}$ different basis states in the address register.
For example, assuming Pauli-$Y$ rotations,
\begin{equation}
R_y(\phi) := e^{-\I Y \phi/2} = 
\begin{bmatrix}
    \cos \nicefrac{\phi}{2} &  -\sin \nicefrac{\phi}{2}  \\
    \sin \nicefrac{\phi}{2} & \phantom{-}\cos \nicefrac{\phi}{2} \\
\end{bmatrix},
\label{eq:ry_mat}
\end{equation}
the unitary matrix corresponding to a $\UCRy$ gate 
with the final qubit as the data qubit
is given by the following block diagonal matrix,
\begin{equation}
\small{
    \UCRy(\angles) =
    \begin{bmatrix}
    R_y(\apt_0) &        & \\
                & \ddots & \\
                &        & R_y(\apt_{2^{n_a}-1})
    \end{bmatrix},   
}
\label{eq:ucry_mat}
\end{equation}
where $\angles = \left[ \apt_0, \ldots, \apt_{2^{n_a}-1} \right]$
is a vector of rotation angles.
It follows that~\cite{Amankwah2022QuantumImages}
\begin{align}
\psifrqi[\myvec{\alpha}] & = \UCRy(\myvec{\alpha}) \, (H^{\otimes n_a} \otimes I) \, \ket{0}^{\otimes (n_a + 1)}, \nonumber \\
& = \sum_i \ket{i} \otimes (\cos \nicefrac{\alpha_i}{2} \ket{0} + \sin \nicefrac{\alpha_i}{2} \ket{1}).
\label{eq:psifrqi}
\end{align}
To recover $\angles$ through projective measurement of $\psifrqi[\angles]$
in the computational basis, we sample from the probability density function (PDF)
$\big|\!\psifrqi[\angles]\!\big|^2 = \big|\!\left[ c_{0}, s_{0}, \ldots, c_{2^{n_a}-1}, s_{2^{n_a}-1} \right]\!\big|^2$,
where $c_i = \cos \nicefrac{\alpha_i}{2}$ and $s_i = \sin \nicefrac{\alpha_i}{2}$.
The input angles $\angles$ can be uniquely recovered by measuring the PDF,
\begin{equation}
\label{eq:frqi-rec}
\aptmeas_{i}=2 \arctan{\sqrt{\frac{|s_{i}|^2}{|c_{i}|^2}}}, \qquad
i \in [2^{n_a}],
\end{equation}
provided that $\alpha_i \in [0, \pi]$.
The input data $\data$ should be rescaled to rotation angles in this restricted range, e.g., $\apt_i = \dpt_i / A$, where $A$ is a scaling factor such that
the angles are mapped to $[0, \pi]$.

A straightforward circuit implementation of the $\UCRy$ gate in \cref{eq:ucry_mat} consists of $2^{n_a}$ fully-controlled $R_y$ gates, where, for $i \in [2^{n_a}]$, the rotation angle is given by $\alpha_i$ and the $n_a$ address qubits are controlled on the state $\ket{i}$.
An optimized circuit implementation existing of a depth-$2^{n_a}$ sequence of two-qubit \cnot\ gates and uncontrolled single-qubit $R_y$ rotations~\cite{Mottonen2004} reduces the quantum resources~\cite{Amankwah2022QuantumImages} at the cost of an increased classical overhead to
solve the linear system
\begin{equation}
\theta_j = \sum_{i} W_{i,j}^{\prime} \alpha_i,  \quad \text{for}\quad i,j \in [2^{n_a}],
\label{eq:walsh}
\end{equation}
for the rotation angles $\myvec{\theta}$.
The angles $\theta_i$ are the parameters that are used in $R_y$ rotations of the compact circuit implementation for the $\UCRy$ gate in \cref{fig:pUCR}(a).
The linear system~\eqref{eq:walsh} is a \emph{Walsh-Hadamard} transformation with Gray ordering, explained in more details in \cref{app:pUCRy}. 
It can be solved efficiently classically in $\bigO(N \log N)$ operations through a \emph{Fast Walsh-Hadamard Transform} (FWHT)~\cite{Amankwah2022QuantumImages}.

In another related work, we proposed FABLE~\cite{Camps2022a}, which leverages compact $\UCRy$ and $\UCR_z$ circuits to generate block-encodings of matrices, a widely used primitive in quantum linear algebra algorithms such as the quantum singular value transformation~\cite{Gilyen2019, Martyn2021}.

Our new data encoding \qcrank\ is an extension of the QPIXL--FRQI and MCRQI angle encodings.
It uses parallel single-qubit rotations and \cnot\ gates acting on disjoint qubit pairs leading to much shorter circuits with a high degree of quantum-parallelism.
Our new \qbart\ basis encoding is a \qcrank\ derivative that generates compact circuits for QIRs which use a basis encoding for the color mapping $\ket{c_i}$ such as NEQR.
We do consider both \qcrank\ and \qbart\ in the broader context of encoding \emph{ordered data} $\data$ in a quantum state following \cref{eq:enc},
where the ordering of $\data$ is imposed by the tensor product of states on the address $\ket{i}$ and data qubits $\ket{c_i}$.
\cref{eq:enc}
can be viewed as a generic case of a vectorized data structure with $\ket{i}$ the index and $\ket{c_i}$ encodes the value of $\dpt_i$, respectively.  
This general quantum \emph{index-value data structure} enables a natural representation and manipulation of different types of ordered data, such as DNA sequences, complex-valued series, 2D images, and time-ordered ECG waveforms, as shown by the experiments presented in the next section and in \cref{app:experiments}.
The experiments leverage  
\qcrank- and \qbart-based quantum algorithms and are executed on either NISQ~\cite{Preskill2018} hardware or noisy simulators.
This work demonstrates that today's NISQ devices can encode and compute on still simplified but real-world problems.

\section{Results}
\label{sec:results}

In this section, we describe the main contributions of this paper, including the optimally scheduled parallel UCR gates, our new encodings \qcrank\ and \qbart, and a series of experiments with various types of data demonstrating their performance on real QPUs.

\subsection{Optimally scheduled parallel UCR gates}
\label{ssec:pUCR}

\begin{figure*}[t]
\centering
\includegraphics[width=\textwidth]{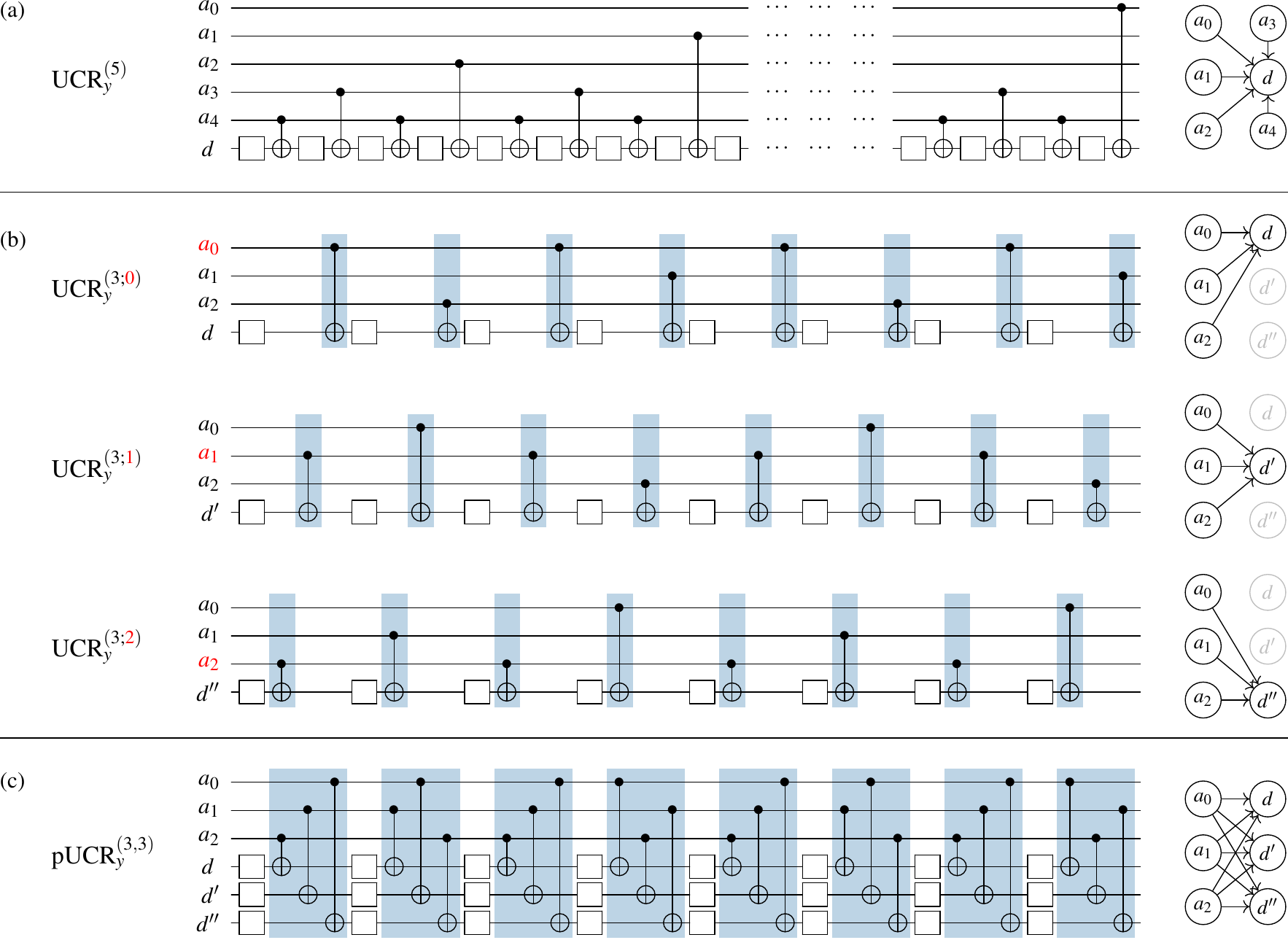}
\caption{Different types of uniformly controlled rotations (UCR) with optimal connectivity graphs for qubits shown on the right. (a) Standard compact circuit implementation for a $\UCRy$ gate that was used in the QPIXL framework~\cite{Amankwah2022QuantumImages} for 5 address and 1 data qubits. Square boxes denote single qubit $R_y$ rotations. (b) All 3 possible realizations of the cyclic permuted UCRs for 3 address and 1 data qubits. (c) Parallel UCR for 3 address and 3 data qubits. The same  3 different permuted $\UCRy$ circuits using the common 3 address qubits and 3 different data qubits can be reordered to an equivalent circuit with the same CX depth as a single $\UCRy$ circuit. Blue rectangles indicate groups of  3 CX gates which can be executed concurrently in the same cycle.}
\label{fig:pUCR}
\end{figure*}

\cref{fig:pUCR}(a) illustrates the standard compact circuit implementation for a $\UCRy$ gate~\cite{Mottonen2004} with  $n_a=5$ address qubits that was used in the QPIXL framework~\cite{Amankwah2022QuantumImages}.
However, the $\UCRy$ circuit implementation is not unique. The positions of the control qubits of the \cnot\ gates can be permuted in a cyclical manner, as shown in \cref{fig:pUCR}(b).
We denote $\UCRy[n_a;s]$ as the circuit implementation of a uniformly controlled $R_y$ rotation with $n_a$ address qubits and \emph{cyclic permutations} $s \in [n_a]$.
We show all 3 possible realizations of cyclic permuted UCRs for 3 address qubits, i.e., $s = 0 \to [0,1,2]$, $s = 1 \to [1,2,0]$, and $s = 2 \to [2,0,1]$.
Note that the different implementations of the $\UCRy$ gate require a slight modification
to the linear system~\eqref{eq:walsh} in order to compute the rotation angles. The angles
can always be computed with an $\bigO(N \log N)$ algorithm.
More details are provided in \cref{app:pUCRy}.

The benefit of the permuted $\UCRy$ gates becomes clear when we combine multiple of them acting on different data qubits but sharing the same address qubits, as shown in ~\cref{fig:pUCR}(c). 
In this case, the 1- and 2-qubits gates expressing the $\UCRy$ circuits with the 3 different permutations acting on 3 data qubits 
can be reordered to an equivalent circuit with the same critical depth as a single $\UCRy$ circuit acting on 1 data qubit. 
This is possible because the single-qubit $R_y$ rotations act on different qubits and groups of  \cnot\ gates are acting on different pairs of qubits.
Consequently, both the $R_y$ and \cnot\ gates mutually commute and can be reordered to allow for concurrent execution. 
We call this a \emph{parallel uniformly controlled rotation} gate or $\pUCRy[n_a, n_d]$ with $n_a$ address qubits and $n_d$ data qubits.
The \cnot-depth of a $\pUCRy[n_a, n_d]$  circuit is 
\begin{equation}
    d_{\cnot} = \ceil{n_d/n_a} \, 2^{n_a}, ~~~ \text{for}~~ n_d,n_a>0.
    \label{eq:cx-depth}
\end{equation}
If $n_d \leq n_a$, then the \cnot\ gates within a cycle act along edges in the bipartite connectivity graph shown on the right of ~\cref{fig:pUCR}(c) that connect disjoint pairs of address and data qubits.
Consequently, these \cnot\ gates
can be executed in parallel on the quantum hardware, significantly shortening the execution time and improving the fidelity.

A $\pUCRy[n_a, n_d](\angles)$ gate implements the block diagonal unitary
\begin{equation}
    \begin{bsmallmatrix}
    R_y(\apt_{0,0}) \otimes \cdots \otimes R_y(\apt_{0,n_d-1}) &        & \\
                & \ddots & \\
                &        & R_y(\apt_{2^{n_a}-1, 0})\otimes \cdots \otimes R_y(\apt_{2^{n_a}-1, n_d-1})
    \end{bsmallmatrix},
\label{eq:pucry_mat}
\end{equation}
with $\angles = \left[ \apt_{i,j} \right]$  a vector of  $n_d \times  2^{n_a} $ rotation angles that encode the data $\data$.

\subsection{QCrank}
\label{sec:qcrank}

\begin{figure}[hbtp]
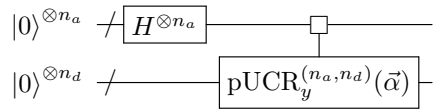

\centering
\(\small
\qquad\qquad
\begin{myqcircuit}
\lstick{\ket{0}^{\otimes n_a}} & {/} \qw & \gate{H^{\otimes n_a}} & \ctrlsq{1} & \qw &&\\
\lstick{\ket{0}^{\otimes n_d}} & {/} \qw & \qw & \gate{\pUCRy^{(n_a,n_d)}(\angles)} & \qw &&
\end{myqcircuit}
\)
\caption{High-level block diagram of the \qcrank\ circuit  encoding  $n_d \times  2^{n_a}$ real values $\angles$ onto the state of $n_a+n_d$ qubits.
}
\label{circ:hlqcrank}
\end{figure}

The quantum-parallel data encoding scheme that we propose in this paper leverages the $\pUCRy$ circuits to generate an encoding.
To this end, we only have to prepend {\it Hadamard} gates acting on the register of  address qubits of the $\pUCRy$ circuit. 
That creates an equal superposition over all addresses as required for \cref{eq:enc}.
\cref{circ:hlqcrank} shows the high-level block diagram of the \qcrank\ circuit. We call our method \qcrank\ as the arrangement of the \cnot\ gates in the $\pUCRy$ circuit diagram in \cref{fig:pUCR}(c) resembles a crankshaft in a combustion engine.

Similar to \cref{eq:psifrqi}, it follows from \cref{eq:pucry_mat} that the \qcrank\ circuit prepares the state:
\begin{align}
\psiqcrank[\angles]
&=
\frac{1}{\sqrt{2^{n_a}}} \sum_{i=0}^{2^{n_a}-1} \ket{i} \otimes \ket{c_{i, 0}} \otimes \ket{c_{i, 1}} \otimes \cdots \otimes \ket{c_{i, n_d-1}}, \nonumber\\
\ket{c_{i, j}} &= \cos(\apt_{i,j}/2) \ket{0} + \sin(\apt_{i, j}/2)\ket{1},
\label{eq:qcrank}
\end{align}
where $j \in [n_d]$.
For a fixed index $i$ and corresponding state $\ket i$ on the address qubits,
the different rotation angles $\apt_{i, 0}, \ldots, \apt_{i, n_d-1}$ are encoded in the product state
$\lvert c_{i, 0} \rangle \cdots \lvert c_{i, n_d-1} \rangle$. This implies that the data recovery process can be decoupled into $n_d$ independent vectors of input parameters $\angles_{:, 0}, \ldots, \angles_{:, n_d-1}$, where $\angles_{:,j}$ denotes the vector  constructed by taking all values with second index equal to $j$.
By tracing out all data qubits except the $j$th, the FRQI state~\cite{Le2011AOperations} corresponding to
input parameter $\angles_{:,j}$ is retrieved. Formally,
\begin{equation}
\rho_\text{FRQI}(\angles_{:, j}) = \trace_{\mathcal{D}_{k: k \neq j}} \rho_{\text{qcrank}}(\angles),
\label{eq:trace}
\end{equation}
where $\mathcal{D}_{k: k \neq j}$ is the Hilbert space of all data qubits except the $j$th,
and $\rho_\text{FRQI}$ and $\rho_{\text{qcrank}}$ are the density matrices defined in the usual manner.
\cref{eq:trace} provides a procedure to reduce the PDF on $n_a+n_d$ qubits  measured from \qcrank\ to $n_d$ PDFs
on $n_a$ qubits for an FRQI encoding, for which we can recover the data through \cref{eq:frqi-rec}.
As such, the state $\psiqcrank[\angles]$ can be used to encode a data set $\data$ of size $2^{n_a} \times n_d$
by mapping and/or  rescaling real values $\dpt_{i,j}$ to \qcrank\ inputs $\apt_{i,j} \in [0, \pi]$.
We note that \qcrank\ uses shared address qubits with different data qubits, naturally resulting in shorter circuit depths as shown in~\cref{fig:pUCR}.
We remark that the \qcrank\ state preparation defined in \cref{eq:qcrank}
is mathematically equivalent to the MCRQI encoding~\cite{Sun2013AnComputers}. 

In an idealized setting,
\qcrank\ allows for a lossless encoding.
Moreover, if a very large number of shots is used, $\datameas$ can be recovered up to an arbitrary precision.
In reality, the performance of
NISQ-era hardware is still severely restricted by gate infidelities, short coherence times, and cross-talk. 
For a NISQ device, we cannot expect that the reconstruction error
decreases monotonically just by increasing the numbers of shots. 
For example, slight under- or over-rotation during the $R_y$ rotations on the data qubits can accumulate and distort the relation between the {\it intended} and {\it achieved } rotation angles.
Moreover, \cnot\ errors are typically an order of magnitude higher and lead to non-local errors.
To compensate for all these effects, 
we introduce a hardware and circuit dependent heuristic function $g(\cdot)$, called {\it adaptive calibration}, which corrects the angles obtained from \cref{eq:frqi-rec} to allow for a near-perfect \qcrank\ decoding on NISQ devices,
\begin{equation}
\alpha^{\text{meas}*}_i = g\left( \aptmeas_i \right), 
\qquad i \in [2^{n_a}].
\label{eq:the-rec-nisq}
\end{equation}
For clarity we distinguish parameters recovered with the heuristic \cref{eq:the-rec-nisq} by adding an asterisk as a superscript.  The construction of  $g(\cdot)$
from calibration measurements
is discussed in more details in \cref{sec:methods}.

For the \qcrank\ experiments on NISQ hardware discussed in this work, we further limit the generality of the input data $\data$ from continuous to discrete.
We take that $\data$ consists of a sequence of values drawn from a discrete set $[0, 1, \ldots, K-1]$. 
We call each possible input value a \emph{symbol} and interpret $K$  as the \emph{max value} or $\ceil{\log_2(K)}$ as the \emph{bit depth} in the case of digitized  sequences, images, or time-series data.
The advantage here is that to recover the discretized data $x^{\text{meas}*}$, we only need to be able to distinguish $K$ different $\apt^{\text{meas}*}$ values that are spaced $\nicefrac{\pi}{K}$ apart.

Despite all of the advantages of the \qcrank\ encoding, we must point out that there are two difficulties in using it for large-scale data processing.
\emph{First}, decoding the data from \qcrank\ relies on accurately measuring the PDF in order to compute the data through~\eqref{eq:frqi-rec}, a problem that clearly scales exponentially 
with the number of address qubits $n_a$.
For example, assuming 8 address qubits and 16 data qubits, \qcrank\ can store $2^8 \times 16 = 2^{12}$ real input values onto a QPU onto 24 qubits. 
To recover all $2^{12}$ real values, we need to accurately measure probabilities for all $2^8=256$ address bit-strings, separately for each data qubit. 
Assuming we aim for an error of 1\%  for  probabilities used in  \cref{eq:frqi-rec}, it would require $\mathcal{O}(10^4)$ shots per bit-string. 
Hence, $\mathcal{O}(10^6)$ shots would be required to recover all stored values with $\mathcal{O}(1\%)$ precision.
This estimate on the number of shots is based purely on sampling error and does not consider the infidelity of the actual quantum hardware, which will further increase the necessary amount of shots. 
Using \qcrank\ can be practical if we plan to recover only a small number of values from the QPU, but even in this case, post-selection on the address bit incurs an exponential overhead.
Furthermore, merely encoding and decoding classical data on a QPU is of limited interest outside of benchmarking and verification purposes.
Consequently, some quantum data processing that condenses the information from the high-dimensional input space to a low-dimensional {\it solution} needs to be applied on the QPU before we read it back classically.
\emph{Second,} it is not trivial to come up with data processing algorithms that act on the angle encoding used in \qcrank\ and return a condensed result.
To overcome these issues, we propose the \qbart\ encoding which uses the NEQR {\it basis encoding}.

\subsection{QBArt}
\label{sec:qbart}

The \emph{Quantum Binary representation Arithmetic} (\qbart) encoding retains the quantum-parallel feature of \qcrank\ while encoding the data in 
the well-studied basis encoding as used in NEQR~\cite{Zhang2013a}.
Formally, \qbart\ generates circuits with identical structure as the \qcrank\ circuit (\cref{circ:hlqcrank}), except that the rotation angles are now restricted to two discrete values $\angles \in \lbrace 0, \pi \rbrace$.
It was noted in Figure 3 and Definition 6 in our previous work~\cite{Amankwah2022QuantumImages} that serial $\UCRy$ gates can be used to prepare an NEQR state. 
\qbart\ instead leverages the compact parallel UCR circuits to efficiently prepare the NEQR state on real QPUs.

\qbart\ offers a lower density of information storage than \qcrank\ because data qubits now hold only superposition of $\lbrace 0,1\rbrace$'s instead of  superposition of real numbers encoded as $R_y$ rotations.
However, the output of \qbart\ is sparse, so decoding requires a far smaller number of shots. 
Furthermore, instead of relying on estimating the PDF and using~\cref{eq:frqi-rec} to recover the data, the observed bit-strings themselves contain the data.
Theoretically, that leads to an exact data value reconstruction using a single observation.
We will experimentally demonstrate that post-processing by {\it majority voting} suppresses the noise artifacts very effectively
for \qbart\ executed on a NISQ hardware.
The number of data qubits used in \qbart\ sets the final precision with which a quantum computation is performed on the data, regardless of the QPU's fidelity. 
Since many data processing
tasks require 8 to 16 bits of precision, this is a manageable overhead in terms of the qubit count, even for existing QPUs.

In the following, we demonstrate that the proposed quantum encodings \qbart\ and \qcrank\ enable today's NISQ devices to encode \emph{and} 
process data stemming from real-world problems, such as DNA sequence matching, processing of time-series data, or 2D images.

\section{Experiments}

In the following, we describe several experiments executed on real hardware provided by Quantinuum, IBMQ, and IonQ, as summarized in \cref{tab:experiments}.

\setlength{\tabcolsep}{3.5pt}
\begin{table}[hbtp]
\centering
{\small
\begin{tabular}{l|cccc|c} 
\toprule
  & exp \#1 & exp \#2 & exp \#3 &exp \#4 & exp \#5\\
\midrule
{\bf Encoding} & \qbart & \qbart & \qbart & \qcrank & \qbart\\
\midrule
{\bf Data type} & \multicolumn{2}{c}{DNA} & Time- & Binary & Integer\\
& \multicolumn{2}{c}{sequence} & series & image& sequence \\
\midrule
{\bf Objective} & DNA & Hamm. & Complex & LBL logo & Hardware \\
& match & weight & conjugate &  I/O & benchmark\\
\midrule
real QPU & H1-1 & H1-1 & H1-1 & H1-1 & diverse$^\dag$\\
addr. qubits & 4 & 4 & 5 & 4 & 2 \\
data qubits & 12 & 3 & 10 & 8 & 4\\
ancillas & - & 1 & - & - & -\\
reset ops & 5 & - & - & - & - \\
\midrule
input (bits) & 192 & 48 & 320 & 384$^\star$ & 16\\
\bottomrule
\end{tabular}\\[5pt]
{\footnotesize $^\dag$QPU hardware provided by Quantinuum, IBMQ, and IonQ.}\\
{\footnotesize $^\star$Assuming 3-bit resolution per real value encoded by \qcrank.}
}
\caption{\qcrank\ and \qbart\ experiments executed on the real hardware.}
\label{tab:experiments} 
\end{table}

\subsection{DNA sequences}

The genetic code of any organism is described as a sequence of {\it codons} that encode specific amino acids. 
A codon consists of three nucleotides. 
Since 4 types of nucleotides exist in nature ($A$, $T$, $G$, $C$), there are 64 different codons, which  
 is equivalent to 6 classical bits of information. 

On a quantum computer, we will use 6 qubits to encode a codon by assigning 2-qubit basis states to the 4 nucleotides,
\begin{align}
    A &= \ket{00}, &
    T &= \ket{01}, &
    G &= \ket{10}, &
    C &= \ket{11},
\end{align}
and constructing  the tensor product of 3 2-qubit states, e.g., $ACT = \ket{001101}$ or $ATG = \ket{000110}$.
In order to compare 2 DNA sequences made of codons, we compute on pairs of codons, requiring 12 qubits in total.
Consequently, the integer values in \cref{eq:enc}, $c_i \in [4096]$, allow the encoding of 2 codons as a quantum state $\ket{a_0 \cdots a_5 \, b_0 \cdots b_5}$, being again a tensor product of the quantum states of 2 codons.

\paragraph{Pattern matching}
In \emph{Experiment \#1}, the inputs are two codon sequences of equal length and the output is a 1-bit sequence of the same length that contains 1 for every position where the codons match and 0 elsewhere.
The additional 6-bit wide output sequence encodes which of nucleotides did not match.

\begin{figure}[h!]
\centering
\(\small
\quad
\begin{myqcircuit}
 \lstick{\ket{0}^{\otimes 4}} &{/}\qw   &\multigate{12}{\rotatebox{90}{\qbart$(~\vec A, \vec B~)$}}  &\qw & \qw   & \qw &  \qw &  \qw   & \qw  &  \qw   &  \qw  &  \qw &  \qw & \qw   & \qw     & \qw  & \qw  &  \meter & \rstick{\text{addr}}\\
  \lstick{\ket{0}_{a_0}} &\qw   &\ghost{Uu}  &\qw &  \targ    & \qw   & \qw & \qw  & \qw   & \qw   & \gate{X} & \ctrl{1} & \qw   & \qw    & \qw  & \qw  & \gate{X}&  \meter & \rstick{p_0}  \\
  \lstick{\ket{0}_{a_1}} &\qw   &\ghost{Uu}  &\qw & \qw &  \targ  & \qw   & \qw  & \qw & \qw  & \gate{X} & \ctrl{5} & \qw   & \qw & \qw   & \qw  & \gate{X}  &  \meter & \rstick{p_1}  \\
  \lstick{\ket{0}_{a_2}} &\qw   &\ghost{Uu}  &\qw & \qw  & \qw  &  \targ  & \qw& \qw & \qw  & \gate{X} & \qw    & \ctrl{1} & \qw    & \qw & \qw   & \gate{X}&  \meter & \rstick{p_2}    \\
  \lstick{\ket{0}_{a_3}} &\qw   &\ghost{Uu}  &\qw & \qw  & \qw  & \qw &  \targ  & \qw  & \qw  & \gate{X} & \qw    & \ctrl{4} & \qw &\qw  & \qw   & \gate{X} &  \meter & \rstick{p_3}   \\
  \lstick{\ket{0}_{a_4}} &\qw   &\ghost{Uu}  &\qw & \qw  & \qw  & \qw & \qw &  \targ  & \qw  &   \gate{X}  & \qw   &\qw   &\ctrl{1} & \qw  & \qw  & \gate{X}&  \meter & \rstick{p_4} \\
  \lstick{\ket{0}_{a_5}} &\qw   &\ghost{Uu}  &\qw & \qw  & \qw  & \qw & \qw  & \qw &  \targ  &  \gate{X}  & \qw   &\qw    &\ctrl{3} & \qw  & \qw & \gate{X}&  \meter & \rstick{p_5} \\
%
  \lstick{\ket{0}_{b_0}} &\qw   &\ghost{Uu}  &\qw & \ctrl{-6}&  \qw  &  \qw & \qw  & \qw   & \qw  &  \push{\ket{0}}  & \targ &  \qw  & \qw   &\ctrl{2} &\qw  &\qw \\ 
  \lstick{\ket{0}_{b_1}} &\qw   &\ghost{Uu}  &\qw &  \qw & \ctrl{-6} & \qw & \qw&  \qw  & \qw & \push{\ket{0}}    & \qw & \targ &  \qw  &\ctrl{2}   &\qw  &\qw  \\
  \lstick{\ket{0}_{b_2}} &\qw   &\ghost{Uu}  &\qw & \qw  & \qw & \ctrl{-6} & \qw & \qw   & \qw  &  \push{\ket{0}}   &  \qw& \qw & \targ &  \qw     &\ctrl{2} &\qw  \\
  \lstick{\ket{0}_{b_3}} &\qw   &\ghost{Uu}  &\qw & \qw  & \qw  & \qw & \ctrl{-6} & \qw    & \qw &  \push{\ket{0}}   &  \qw& \qw &  \qw& \targ  & \ctrl{1} &\qw  \\
  \lstick{\ket{0}_{b_4}} &\qw   &\ghost{Uu}  &\qw & \qw  & \qw  & \qw  & \qw  & \ctrl{-6} &   \qw  & \push{\ket{0}}  & \qw&   \qw   & \qw & \qw  &\targ &\qw  &     \meter & \rstick{m_0} \\
  \lstick{\ket{0}_{b_5}} &\qw   &\ghost{Uu}  &\qw & \qw  & \qw  & \qw  & \qw  & \qw  & \ctrl{-6} & \\
{\gategroup{1}{5}{13}{10}{1.em}{--}}
{\gategroup{1}{12}{12}{16}{0.8em}{--}}
    &  & & & &&\text{XOR}(\vec A, \vec B)   & & & &&&&\text{AND}(\bar p_0,...,\bar p_5)
\end{myqcircuit}
\)\\[10pt]
\qquad\quad\includegraphics[scale=0.6]{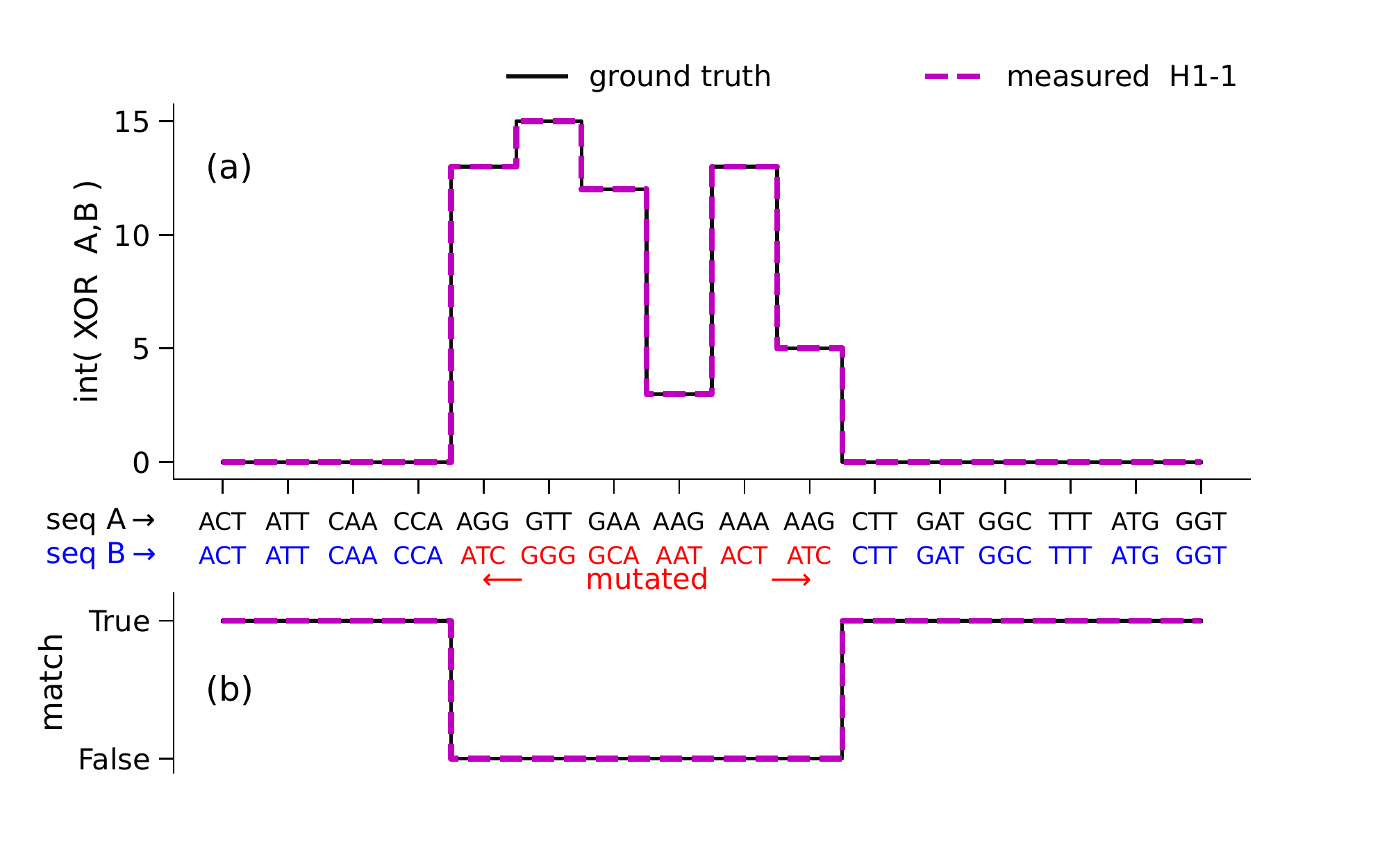}
\vspace{-25pt}
\caption{(Top) Quantum circuit computing the match between two 16 elements long input sequence of 6-bit values A and B,  encoded on qubits $a_i$ and  $b_i$,  respectively. The \qbart\ unitary is  preparing the initial state. The 6 CX gates  compute XOR between bits of sequences element.  The following 5 nested Toffoli gates set the output qubit $m_0$ to state $\ket{1}$ if both 6-bit input pairs match. The intermediate XOR output result  is  also measured on qubits $ p_i$.
(Bottom) Results obtained by the DNA sequence matching  executed on Quantinuum H1-1 QPU. The algorithm correctly detects the differences between the 6 codons in positions from 5 to 10, marked in red. (a) 6-bit XOR$(\vec A, \vec B)$ output sequence and (b) measured match-bit, both follow the ground truth.}
\label{fig:dna-parity}
\end{figure}

The \qbart\ circuit (\cref{fig:dna-parity}) implementing {\it Experiment \#1} requires a total of 16 qubits, 4 of which are used as address qubits and the 12 data qubits encode the codons of both sequences using 6 bits each.
To save on quantum resources, we `recycle' 5 qubits  in the middle of the circuit by applying a {\it reset} gate, which is available on most of QPUs. The total circuit depth is of 68 CX-gates, of which 48 are needed by \qbart\ itself, following \cref{eq:cx-depth}.

We perform this DNA matching experiment on the real 20-qubit trapped-ion QPU from Quantinuum, H1-1. 
The input sequence A is a random snippet from the COVID-19 genome strain~\cite{Sah2020}. 
Sequence B is a copy of A, but the 6 codons
in positions 5 to 10 are randomly altered, as shown in~\cref{fig:dna-parity} (bottom).
At the expense of 600 shots and using the majority voting technique, we achieve an exact result for all 16 codon pairs. 
As shown in~\cref{fig:dna-parity}(a), the 6 XOR bits  $p_i$  are all 0 when the two codons match, and some of them are non-zero otherwise, following the ground-truth. The bit $m_0$, indicating a match, is also correctly computed, as shown in \cref{fig:dna-parity}(b).

\paragraph{Hamming weight computation}

The \emph{Hamming distance} between two bit-strings tells us in how many places they differ from one another; hence, when applied to codon sequences, it is an essential tool for studying the evolution of the genetic code. 
The \emph{Hamming weight} of a binary string is defined as the number of bits set to 1.
In {\it Experiment \#1} we have already computed the binary XOR value between the two codons, stored at $p_i$.
We will now pass them to the Hamming weight algorithm to compute the desired Hamming distance.

\begin{figure}[hbtp]
\centering
\(\small
\begin{myqcircuit}
 \lstick{\ket{0}^{\otimes 4}} &{/}\qw   &\multigate{3}{\qbart}  &\qw & \qw  & \qw &  \qw  & \qw     &  \meter & \rstick{\text{addr}}\\
  \lstick{\ket{0}_{p_0}} &\qw   &\ghost{\qbart}  &\qw   & \qw \barrier[0em]{3}  & \qw  &\ctrl{2} &\ctrl{2} &\qw   \\
  \lstick{\ket{0}_{p_1}} &\qw   &\ghost{\qbart}   &\ctrl{1} &\ctrl{1} &  \qw  & \qw  & \qw  & \qw \\
  \lstick{\ket{0}_{p_2}} &\qw   &\ghost{\qbart}  &\ctrl{1} & \targ & \qw   &\ctrl{1} & \targ & \meter &  \rstick{s_0} \\
  \lstick{\ket{0}}  &\qw    &\qw   & \targ   & \qw  &\qw   & \targ   &\qw & \meter &  \rstick{s_1}  \\
   &  &  & ~~~~~~p_1+p_2  & &&~~~~~~~~+p_0& \\
\end{myqcircuit}
\)\\[15pt]
\includegraphics[scale=0.6]{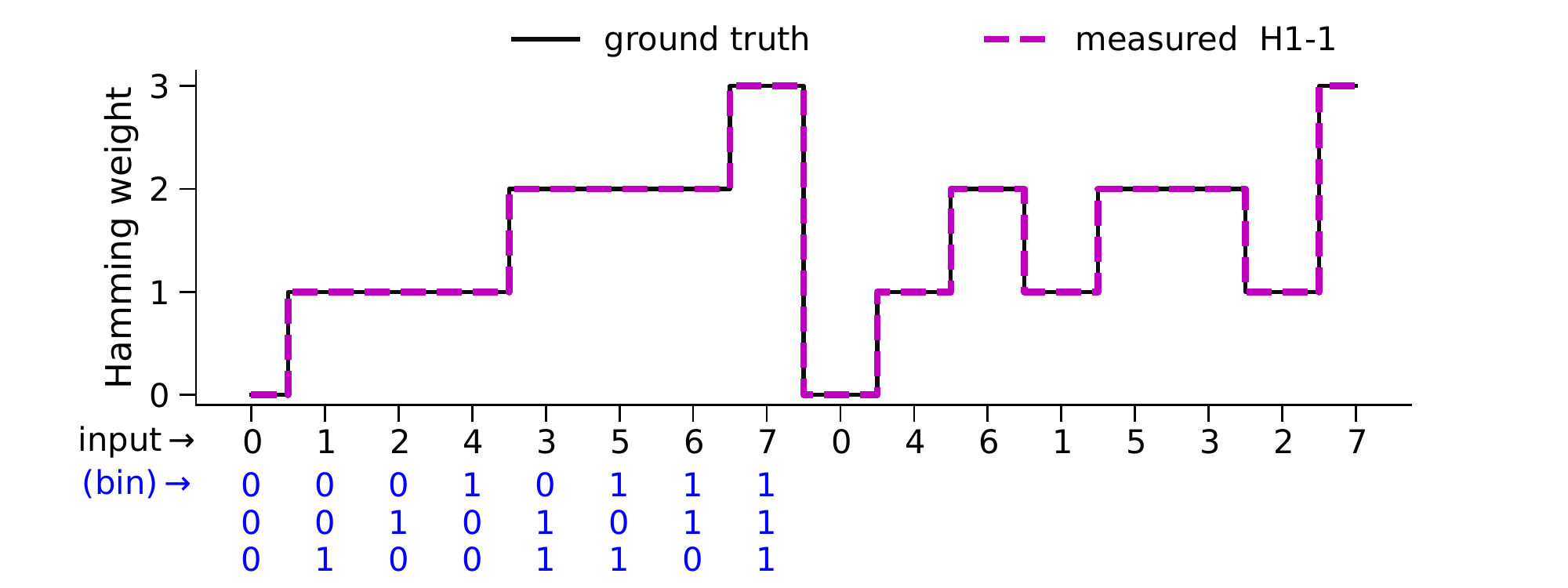}
\caption{(Top) Quantum circuit computing the Hamming weights (HW) for a sequence of 16 3-bit integers.  \qbart\ unitary  encodes the input sequence   on qubits $p_i$. The following 1st sub-ciruit computes the partial HW for inputs $p_1,p_2$. The 2nd one adds the value of $p_0$ to the partial HW stored binary on qubits $s_0,s_1$. The final HWs are measured on qubits $s_0,s_1$, for all addresses. 
(Bottom) Results of Hamming weight computation for a sequence of 16 3-bit integers, executed on Quantinuum H1-1.}
\label{fig:qbart-hammW}
\end{figure}

{\it Experiment \#2} computes the Hamming weights for a 3-bit sequence of length 16. 
We use \qbart\ with 4 address qubits and 3 data qubits to encode the input, and 
1 {\it ancilla} qubit. 
The complete \qbart\ circuit (\cref{fig:qbart-hammW}) uses 8 qubits, has CX-depth of 28 cycles on an {\it all-to-all} connected QPU. 
The experimental results from H1-1 for a pseudo-random input sequence and using 300 shots agree with the ground-truth \textit{exactly}, as shown in \cref{fig:qbart-hammW}.

We recognize that the quantum circuit in \cref{fig:qbart-hammW} does only `half' of the job, since it reduces only 3 inputs  $p_0,p_1,p_2$ to 2 output bits $s_0,s_1$. However, with enough resources, we can add a second copy of the same circuit acting in parallel on the qubits $p_3,p_4,p_5$ from the circuit shown in~\cref{fig:dna-parity} and compute the missing 2nd Hamming weight, to be stored on qubits $s_2,s_3$. Then, we can apply a binary adder on two 2-bit inputs, using one of the known quantum circuits~\cite{Cuccaro2004}, to obtain the full Hamming distance between the 2 codons.

\subsection{Complex conjugate}

\begin{figure}[hbtp]
\centering
\(\small
\begin{myqcircuit}
 \lstick{\ket{0}^{\otimes 5}} &{/}\qw   &\multigate{2}{\qbart(\vec A,\vec B)}  &{/}\qw &  \qw &  \meter & \rstick{t~\text{[addr]}} \\
 \lstick{\ket{0}^{\otimes 5}} &{/}\qw   &\ghost{\qbart(A,B)}  &{/}\qw & \qw &  \meter & \rstick{\phantom{-}A_t } \\
\lstick{\ket{0}^{\otimes 5}} &  {/} \qw &\ghost{\qbart(A,B)}   &{/}\qw  \qw & \gate{X}\qw & \meter & \rstick{-B_t }\\
\end{myqcircuit}
\)\\
\includegraphics[width=\textwidth]{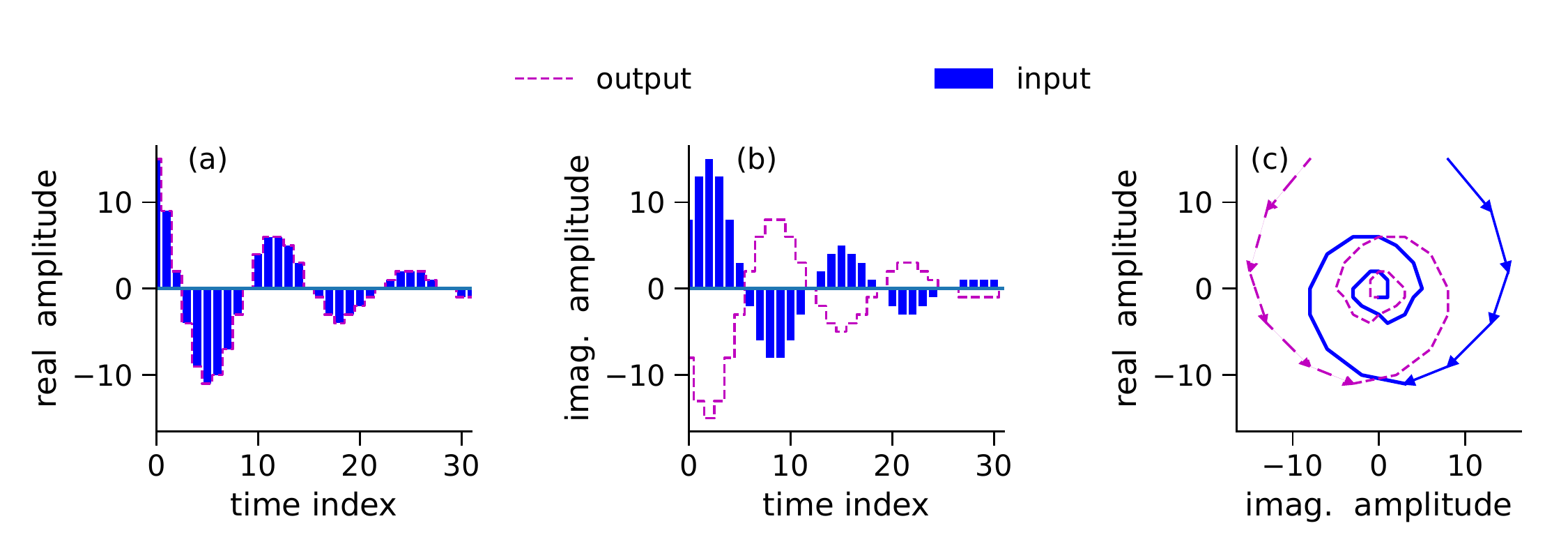}
\caption{(Top) \qbart\ circuit for \textit{experiment \#3} computing the complex conjugate of the input sequence $(A_t,B_t) \rightarrow (A_t,-B_t)$. The values of $\vec A$ and -$\vec B$ are retrieved as {\it signed int} at the addresses  encoded as {\it unsigned int}.
(Bottom) Results of the complex conjugate on complex-valued time-series obtained on the H1-1 QPU are presented in magenta. The input is shown in blue.  (a) and (b) show real and  imaginary components of the pendulum amplitude. (c) depicts its trajectory as function of time. The conjugation operation inverts the sign of the imaginary component.}
\label{fig:qbart-exp8}
\end{figure}

The time evolution of an attenuated pendulum is described by \cref{eq:csin}. 
The real and imaginary components of the complex valued amplitude $C(t)$ are denoted as $A(t)$ and $B(t)$, respectively. They are plotted independently and as a parametric trajectory in blue on 3 panels in \cref{fig:qbart-exp8}.
The purpose of \textit{experiment \#3} is to store the time-series $C_t$ onto the QPU, compute the complex conjugate of the amplitude, $C_t^*$, and recover the resulting new time-series through measurement.
\begin{align}
    C_t &= a \exp{\left[ (b +j c) t +c\right]},    ~~~ C_t \in \mathbb{C},~~~~\text{for}~~~ t \in [0,\ldots,31]\nonumber\\
     A_t &= \real(C_t), \label{eq:csin} \\
     B_t &= \imag(C_t), \nonumber
\end{align}
where $j=\sqrt{-1}$ and the real parameters $a,b,c,d$ are conveniently chosen to match the initial condition $\abs{A_0},\abs{B_0} \simeq 2^4 $.

We use the  {\it signed integer} 5-bit representation for both components $A_t$ and $B_t$, stacked as a single 10-bit input for the \qbart\ circuit (\cref{fig:qbart-exp8}). 
Next, we invert the sign of the $B_t$-data and perform the measurement.
This circuit uses 5 address qubits and 10 data qubits, it has a CX-depth of 64. 
Only one cycle of $X$ gates is needed to invert all 32 $B_t$ values stored in the Hilbert space. 
The output $-B_t$ values are {\it 1's complementary} and need to be decoded as such classically, back to {\it signed} integers.
This method of computing 1's complementary was proposed in the original NEQR paper~\cite{Zhang2013a}.
However, to the best of our knowledge, we present the first practical demonstration of such computation using a real QPU. 
To recover all 32 10-bit values {\it exactly}, the  H1-1 requires $10^3$ shots.

\subsection{2D image}

Any multi-dimensional indexed dataset can always be enumerated as a 1-dimensional sequence. Therefore, our proposed quantum data encodings can directly encode 2D images as well.

\begin{figure}[hbtp]
\centerline{\includegraphics[width=0.75\textwidth]{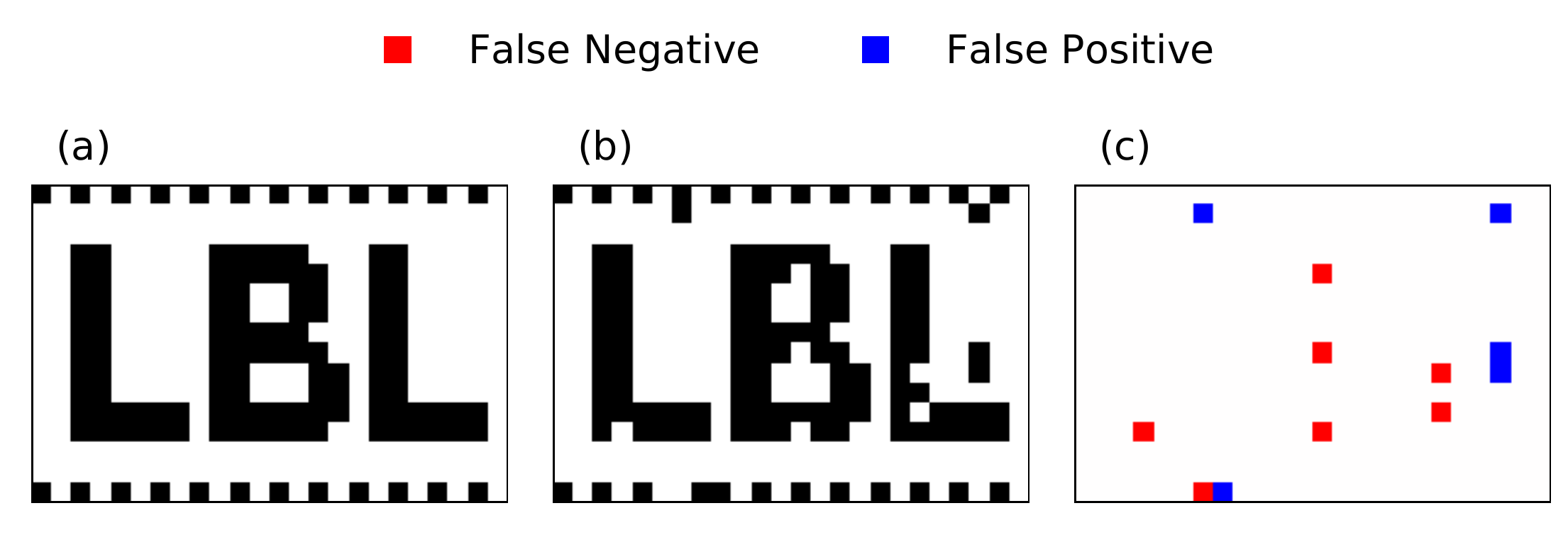}}
\caption{Demonstration of recovery of a black and white 384 pixels image using \qcrank\  \textit{experiment \#4} executed on the Quantinuum H1-1 real QPU. a) ground truth image, b) recovered image has 97\% of correct pixels, c) residual showing the locations of 12 incorrect pixels.}
\label{fig:lbl-logo3}
\end{figure}

{\it Experiment \#4} demonstrates the \qcrank\ encoding of the black and white image of size 384 bits shown in \cref{fig:lbl-logo3}(a).
The recovered image from the Quantinuum H1-1 QPU is shown in \cref{fig:lbl-logo3}(b).
At the expense of  $7000$  shots, \qcrank\ recovers 97\% of the pixels correctly.
This experiment shows that we can store, today, a non-trivial sized image on a 12 qubit system. 
Moreover, an image encoded with \qbart\ could potentially be manipulated by a quantum algorithm, such as filtering~\cite{Jiang2019}.

\subsection{QPUs benchmarking}
\label{sec:QPUs-benchmarking}

A \qbart\ encoding generates the optimal circuit for QPUs on which a bipartite qubit connectivity is naturally available.
The trapped ions QPUs natively allow for such connectivity.
However, for the transmon-based QPUs from IBMQ with {\it heavy-hexagonal} connectivity, the transpiled circuits become a few times deeper due to the inevitable swap operations. 
To compare the performance of diverse types of QPUs,
we reduce the input size to \qbart, leading to a shorter circuit. In this regime, many types of QPUs have a chance to deliver acceptable results.

\qbart\ \textit{Experiment \#5} encodes a sequence of 4 random 4-bit strings with \qbart\ and requires only 2 address and 4 data qubits. Before the transpilation, the circuit depth is 8 CX-cycles. 
We execute this experiment on real QPUs from Quantinuum, IonQ, and IBMQ.
For reference, we also run the same experiment on the ideal \texttt{Qiskit} simulator. \cref{tab:ibmq-hw} in \cref{app:QPU-benchmarks} summarizes the basic characteristics of all benchmarked beckends.

\begin{figure}[b!]
\centerline{\includegraphics[width=1\linewidth]{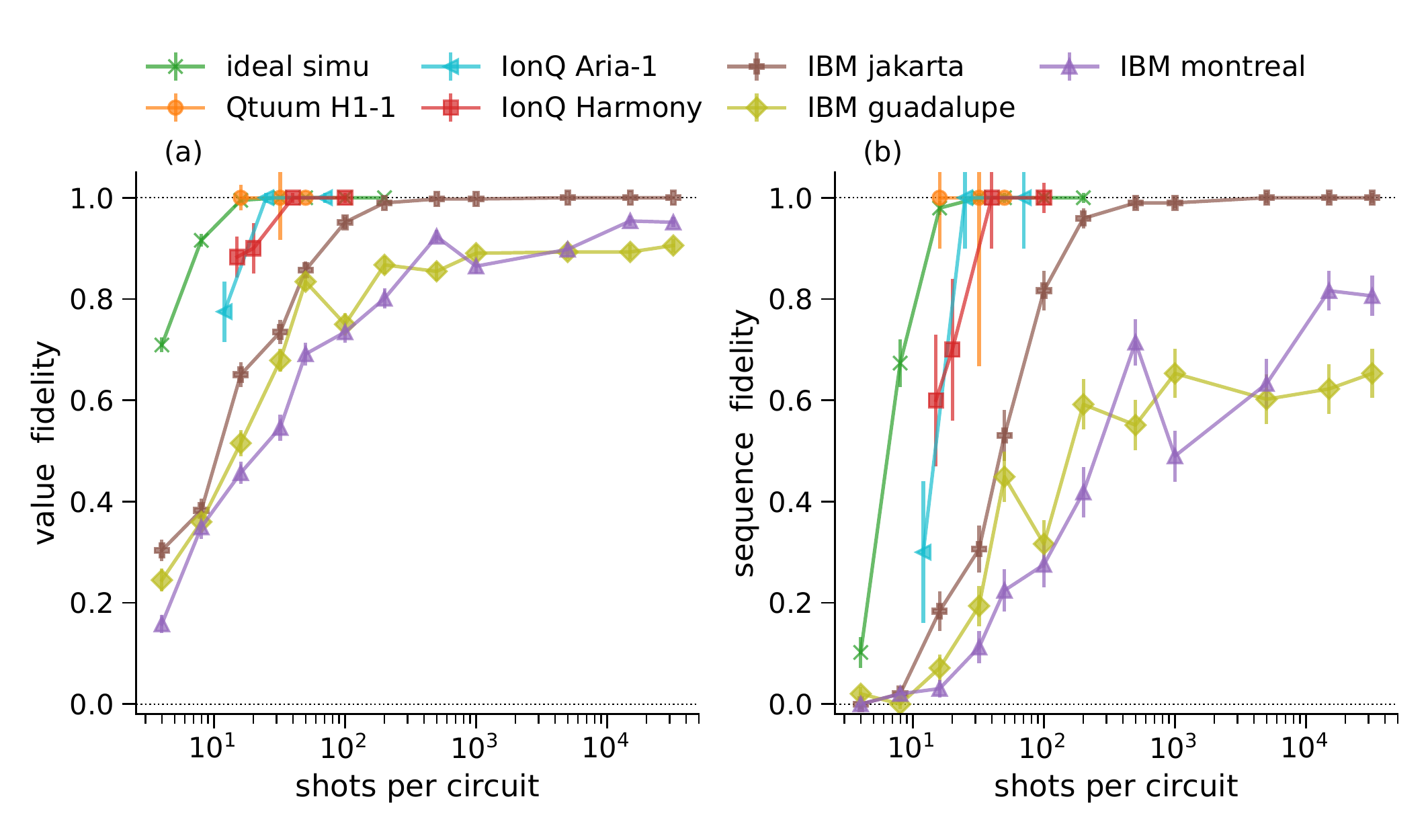}}
\caption{Reconstruction fidelity for \qbart\ experiment \#5 while encoding 4 4-bit integers on 6 qubits, executed on real QPUs provided by  Quantinuum, IonQ, and IBMQ. (a) reconstructed value fidelity (b) reconstructed sequence fidelity. The last H1-1 measurement is for only one input sequence, so no statistical error is presented. }
\label{fig:qbart-exp6}
\end{figure}

The results of \textit{experiment \#5} executed on investigated backends are shown in \cref{fig:qbart-exp6}. We compare shot dependence of two metrics pertaining to (1) the fidelity of individual values in the sequence, and (2) the whole sequence being recovered correctly, both defined in \cref{sec:methods}.
The trapped-ion QPUs require only 100 shots to recover the full sequence and significantly outperform the transmon-based QPUs. 
The two decisive factors are differences in fidelity of entangling gates and versatility of the native connectivity. 
Only on the IBM {\it jakarta}, which has a relatively low CX-gate error, we are able to achieve 100\% sequence fidelity but at the expense of more than 2000 shots.
It requires a 100 times more shots compared to an ideal noisefree QPU.
Here, we also tested the impact of the {\it Dynamical Decoupling} (DD) transpiler pass for IBMQ QPU. The fidelity of the \qbart\ circuit either improves or degrades, depending on a particular QPU; there was no discernible pattern.

\subsection{Simulations}

In addition to the five real hardware experiments described above, we explore the robustness and versatility of our proposed encodings with additional simulated experiments presented in \cref{app:experiments}. In particular, we demonstrate the feasibility of storing an arbitrary waveform on a QPU. We generate a synthetic electrocardiogram (ECG) time-series of length 64, digitize it with the 6-bit resolution, and encode it using \qbart. 
We also study the dynamic range and recovery fidelity for \qcrank\ as a function of the simulated noise and the number of shots.

\section{Discussion}

This work presents two major contributions in the area of quantum data encoding and analysis. 
First, the parallel uniformly controlled rotation circuits (pUCR) enable the storage of a larger input using a shallow circuit because it leverages concurrent execution of elementary
quantum gates on the address and data qubits. 
That leads to compact circuits well-suited for QPUs with a high degree of connectivity, such as ion traps. 
We propose two data encoding methods that make use of pUCR circuits: \qcrank\ which encodes continuous data, and \qbart\ which encodes discrete data in binary representation.
Second, using both \qcrank\ and \qbart, we present an extensive collection of experiments conducted on different real QPUs that demonstrate successful data encoding and analysis at a considerably larger scale than achieved in previous studies. We also develop two error mitigation strategies for \qcrank\ and \qbart\ respectively, to correct the noisy hardware results.

Our experiments show that the Quantinuum H1-1 QPU can reliably prepare a \qcrank\ state that encodes $\bigO(400)$ black-and-white pixels on 12 qubits. 
We introduce an adaptive calibration routine to compensate for the hardware noise and achieve a 97\% recovery fidelity using 7000 shots.
Our experiments with \qbart\ show that the H1-1 QPU can, with near-perfect fidelity, (1) simultaneously encode two DNA sequences of 16 codons stored in 6 bits and compute the positions where the sequences are mismatched, (2) compute the Hamming weight of a sequence of 16 3-bit integers, and (3) compute the complex conjugate of a sequence of 32 complex values with real and imaginary parts both encoded with bit-depth 5. 
We successfully use a majority voting technique to reliably identify the correct results from the measured bit strings. Finally, we report the results of a benchmark comparing the recovered value and sequence fidelity for a \qbart\ encoding with 2 address and 4 data qubits on Quantinuum, IonQ, and IBM QPUs. This experiment highlights the superiority of ion trap QPUs over superconducting QPUs for preparing a \qbart\ state. This is partially attributed to the qubit topology: additional swap gates are required to run a pUCR circuit on superconducting QPUs.

For a fixed number of qubits, the angle encoding used in \qcrank\ and FRQI~\cite{Le2011b} is able to store more data than a binary encoding used in \qbart\ and NEQR~\cite{Zhang2013a}. However, data processing of angle encodings is considerably more difficult compared to binary encodings where classical binary logic operations can be efficiently converted to reversible quantum operations~\cite{NC2010}. Additionally, our experiments show that binary data
can be recovered with greater fidelity and using an order of magnitude fewer shots compared to data stored in an angle encoding.

One exciting topic for future study that our results hint at is the potential of the compact, parallel \qcrank\ circuits in the context of hybrid algorithms such
as Variational Quantum Algorithms (VQA)~\cite{Cerezo2021-jl} or Quantum Machine Learning (QML)~\cite{biamonte2017quantum} tasks. In this context, the rotation angles in the \qcrank\ circuit are considered free parameters that are variationally optimized in a quantum-classical hybrid iteration  where the cost function is evaluated on the QPU. 

\section{Methods}
\label{sec:methods}

\subsection{Metrics of fidelity}
\label{sec:fidelity-metrics}

We define three metrics to characterize the quality of the recovered results using \qcrank\ and \qbart\ encodings on noisy QPUs:
\begin{itemize}
    \item {\bf Dynamic range } ($D_r$) is defined as the distance between the expectation values of the reconstructed angle $\alpha^{meas}$ (\cref{eq:frqi-rec}) for the first and last symbol. It is applicable only for \qcrank\ with input quantized into $K$ symbols $a_0,\ldots,a_{K-1}$
\begin{equation}
    \label{eq:dynr}
    D_r = \frac{\mathbb{E} \big( \alpha^{meas}  ( a_{K-1}) \big) - \mathbb{E} \big( \alpha^{meas} ( a_{0}) \big)}{\alpha( a_{K-1}) - \alpha( a_{0})}.
\end{equation}
The domain of $D_r$ is $[0,1]$, where 1 corresponds to a perfect result and 0 means that pure noise is measured.

\item {\bf Recovered value fidelity} (RVF) is defined as the probability to recover the correct symbol at given position in the sequence, averaged over the sequence.
 
\item{ {\bf Recovered sequence fidelity} (RSF) is defined as the probability of all recovered values in a sequence being correct. 
In the simplest case, $\text{RSF}\sim (\text{RVF})^N$, with $N$ the length of the sequence. }
 
\end{itemize}

\subsection{Adaptive calibration for QCrank}

Based on experiments using a noisy simulator, we observed that the dynamic range ($D_r$) is reduced with increasing gate infidelity, which leads to incorrect symbol reconstruction, as shown in \cref{fig:sim-ang} in \cref{app:experiments}.
Therefore, we developed an {\it adaptive calibration} method to compensate for these distortions. 
For each input symbol, $a_i$, we measure the average reconstructed output $\mathbb{E} (\alpha^{meas}_i$). 
We then define a look-up table with $K-1$ thresholds $\tau_i$,
\begin{eqnarray}
\label{eq:thr-rec}
\tau_i= \frac{ \mathbb{E} (\alpha^{meas}_i) +  \mathbb{E} (\alpha^{meas}_{i+1})}{2},
\end{eqnarray}
set halfway between those $K$ averages, shown as horizontal dotted lines in \cref{fig:sim-ang} in \cref{app:experiments}.
Next, we reanalyze the same experiment and assign the discrete reconstructed symbols using function $g(.), ~g: \alpha^{meas} \rightarrow a^{meas*}$ defined by the following look-up table:
\begin{equation}
    g(\alpha^{meas})= 
\begin{cases}
    a_0,& \text{if~ } \alpha^{meas} < \tau_0\\
    a_{K-1},& \text{if~ } \alpha^{meas} \geq \tau_{m-2}\\    
    a_j,  & \text{if~ }   \alpha^{meas} \in [ \tau_{j},\tau_{j+1}).
\end{cases}
\label{eq:g-alpha-meas}
\end{equation}

\subsection{Majority voting  for QBArt}

For quantum circuit measurements subject to a small noise level, no more than a few of the measured bits for \qbart\ will be incorrect\footnote{The probability for 2 corrupted bits is smaller than for 1 bit.}, resulting in either corrupted address or corrupted data bits.
To compensate for this uncorrelated noise, for each measured address, we choose the reconstructed data to be the most probable value (MPV) of all the data sub-strings collected from the measured bit-strings.
This procedure is equivalent to majority voting on the data bit-strings. 
As the errors on the address bit-strings manifest in the same way, the MPV selection suppresses it as well.
If the values in the sequence are very repetitive, the address error does not matter.

\subsection{Shots count requirement for QBArt}

The output can be accurately recovered for \qbart\, but with a probability that depends on the number of used shots and on the length of the sequence. 
The fundamental notion is the minimal number of appearances of each address sub-string ($M_{min}$) during multi-shot measurements. 
On average, each \qbart\ address is measured with the same probability. 
The Poisson distribution $f(x,\lambda)$ governs the number of appearances of an address.
The {\it lower cumulative distribution}, $P(x,\lambda)$, describes the probability of detecting not more than $x$ occurrences given the average $\lambda$:
\begin{eqnarray}
\label{eq:poiss}
P(x,\lambda)=\Sum^x_{t=0} f(t,\lambda)
\end{eqnarray}
In our case, $\lambda$ is the ratio of the number of shots per circuit to the number of \qbart\ addresses. 
For an ideal QPU, we want each address to appear at least once ($M_{min}=1$), which will happen with probability $1-P(0,\lambda)$. 
For a NISQ device, we will need $M_{min}>1$ to allow for a sufficient number of appearances of the data-bits string at a given address, such that more than once the measured data bit-string is the correct one and the majority voting method selects this bit-string.

It is possible to state the inverse case, i.e. the hardware agnostic problem, as follows.
How many shots per address, $\lambda$, are required to achieve some value of $M_{min}$, while accepting some probability of failure per address, $F_{addr}$? 

\begin{figure}[hbtp]
\centerline{\includegraphics[width=0.75\textwidth]{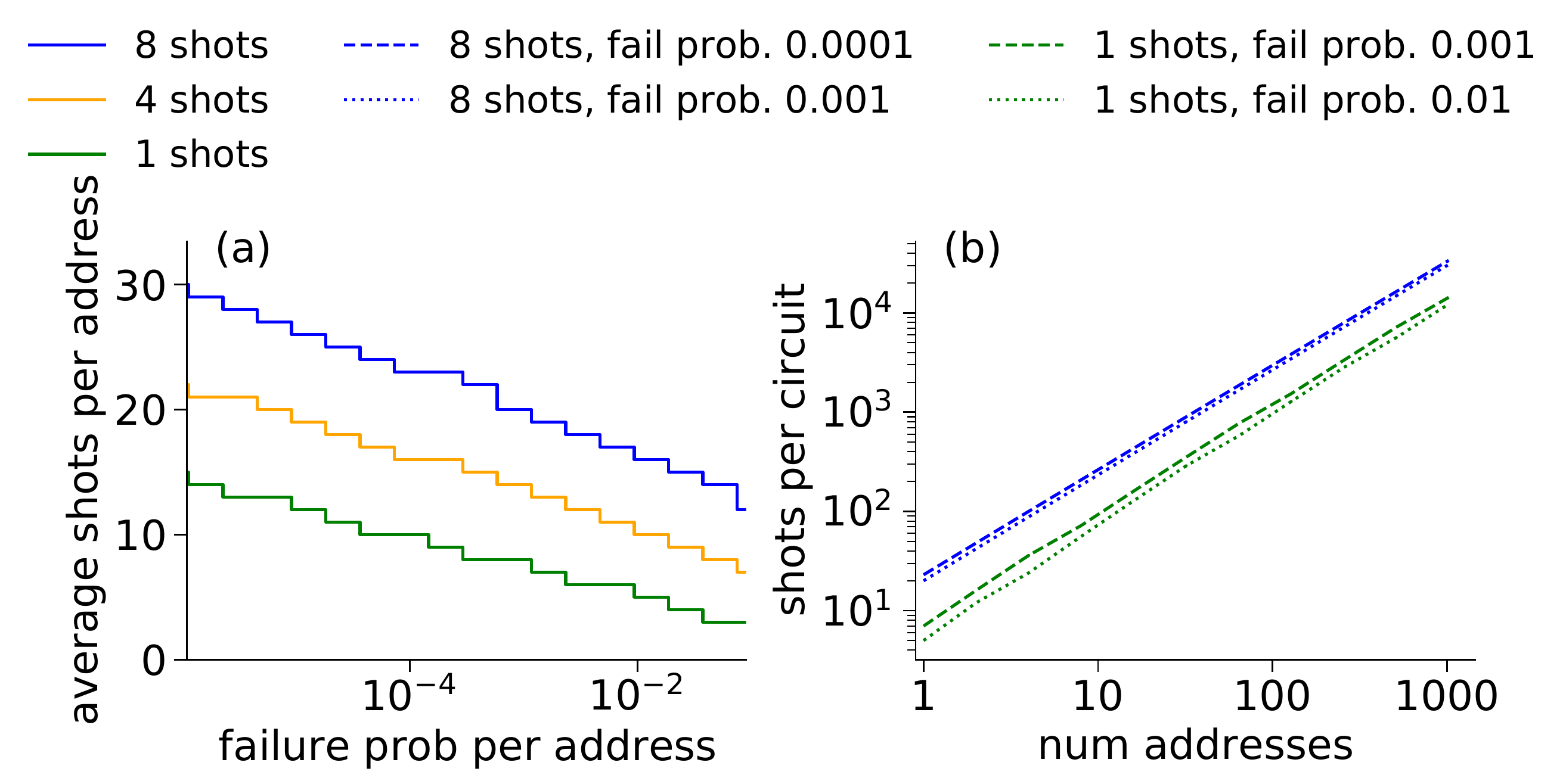}}
\caption{Relationship between the necessary number of shots and the number of \qbart\ addresses. (a) average shots per address sufficient for the 3 choices of minimal number of shots to be guaranteed with probability above 99.9\%. (b) total shots per circuit for several choices of minimal number of shots per address and confidence levels. }
\label{fig:shots}
\end{figure}

\cref{fig:shots}(a) shows analytical results of $\lambda(F_{addr};M_{min})$ for 3 choices of $M_{min}$. 
There is a weak penalty for requiring a larger $M_{min}$. 
In the case of \qcrank, we want the whole data sequence, meaning the values at all addresses, to be reconstructed correctly. 
At the first order, the probability of seeing less than $M_{min}$ appearances ($F_{circ}$) in any of $L$ addresses  equals $L\cdot F_{addr}$. 
\cref{fig:shots}(b) shows the necessary number of shots per \qbart\ circuit as a function of the number of addresses for selected pairs of $(M_{min},F_{circ})$.
The $M_{min}=1$ results relate to an ideal QPU.
It shows that we need only $350$ shots per \qbart\ circuit with 32 addresses to obtain the correct answer with the probability of 99.9\%, regardless of the number of data qubits.
For a NISQ device with H1-1 QPU fidelity level, we may target $M_{min}=8$, which requires $800$ shots instead. 
The dependence of the total number of shots on the failure probability is rather weak.

\subsection{Classical circuit for DNA matching}

\begin{figure}[hbtp]
\centering
 \begin{circuitikz}[scale=0.5 ]
 \draw 
  (1,9) node[xnor port,scale=0.5] (x0) {}
  (1,7) node[xnor port,scale=0.5] (x1) {}
  (1,5) node[xnor port,scale=0.5] (x2) {}
  (1,3) node[xnor port,scale=0.5] (x3) {}
  (1,1) node[xnor port,scale=0.5] (x4) {}
  (1,-1) node[xnor port,scale=0.5] (x5) {}

   (3,8) node[and port,scale=0.5] (a0) {}
   (3,4) node[and port,scale=0.5] (a1) {}
   (5,6.) node[and port,scale=0.5] (a2) {}
   (6.9,-2) node[and port,scale=0.5] (a3) {}
   (3,0.) node[and port,scale=0.5] (a4) {}
   (8.3,9) node[anchor=east] (p0) {$p_0$}
   (8.3,7) node[anchor=east] (p1) {$p_1$}
   (8.3,5) node[anchor=east] (p2) {$p_2$}
   (8.3,3) node[anchor=east] (p3) {$p_3$}
   (8.3,1) node[anchor=east] (p4) {$p_4$}
   (8.3,-1) node[anchor=east] (p5) {$p_5$}
   (7.8,-2) node[ ] (M){$m_0$}
 
  (x0.in 1) node[anchor=east] {$a_0$}
  (x0.in 2) node[anchor=east] {$b_0$}
  (x1.in 1) node[anchor=east] {$a_1$}
  (x1.in 2) node[anchor=east] {$b_1$}
  (x2.in 1) node[anchor=east] {$a_2$}
  (x2.in 2) node[anchor=east] {$b_2$}
  (x3.in 1) node[anchor=east] {$a_3$}
  (x3.in 2) node[anchor=east] {$b_3$}
  (x4.in 1) node[anchor=east] {$a_4$}
  (x4.in 2) node[anchor=east] {$b_4$}
  (x5.in 1) node[anchor=east] {$a_5$}
  (x5.in 2) node[anchor=east] {$b_5$}
  
  (a0.in 1) -| (x0.out)  to [short, -] (p0) 
  (a0.in 2) -| (x1.out)  to [short, -] (p1)
  (a2.in 1) -| (a0.out) 
  (a1.in 1) -| (x2.out)  to [short, -] (p2) 
  (a1.in 2) -| (x3.out)  to [short, -] (p3)
   (a2.in 2) -| (a1.out) 
   (a3.in 1) -| (a2.out)
   (a3.in 2) -| (a4.out)
   (a4.in 1) -| (x4.out)  to [short, -] (p4) 
  (a4.in 2) -| (x5.out)  to [short, -] (p5)

 ; 
  \node at (7.,9) [ocirc]{};
   \node at (7.,7) [ocirc]{};
    \node at (7.,5) [ocirc]{};
     \node at (7.,3) [ocirc]{};
      \node at (7.,1) [ocirc]{};
        \node at (7.,-1) [ocirc]{};
      
 \end{circuitikz}
\caption{ Classical circuit using 6 XNOR, 5 AND, and 6 NOT gates setting bit $m_0$ to true  if  the two 6-bit input registers $a_0,\ldots,a_5$ and $b_0,\ldots,b_5$ are equal.}
\label{fig:class-fpga1}
\end{figure}
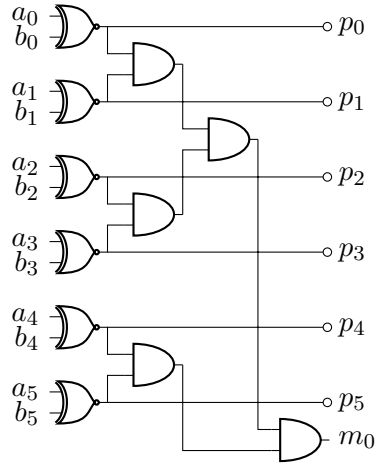

DNA is a sequence of {\it codons} consisting of three {\it nucleotides}.
Given that 4 types of nucleotides exist, the {\it base 64} codon requires 6 classical bits to be encoded. 
Therefore, a reference classical circuit comparing two codons (\cref{fig:class-fpga1}) must have 12 input bits, labeled $a_0,...,a_5, b_0,...,b_5$. 
The first 6 {\it XNOR} gates return 1 if there is a match between their two input bits. The following 5 {\it AND} gates aggregate this information to a single bit $m_0$, set to 1 if all 6 pairs of inputs match. 
The intermediate output of XOR is accessible via bits $p_0,...,p_5$. The $p_i$  bits can be used as input to the following Hamming weight circuit (not shown) producing the Hamming distance between the two input codons.

For the binary encoding of quantum data there is a correspondence between a classical {\it XOR} gate and a quantum {\it CX} gate.
Similarly, a classical {\it AND} gate maps to a quantum {\it Toffoli} gate. 
We exploit this correspondence to construct the quantum circuit in \cref{fig:dna-parity} with almost identical topology as the classical one in \cref{fig:class-fpga1}.

\subsection{Classical circuit for Hamming weight}

\begin{figure}[hbtp]
\centering
\begin{minipage}{0.24\textwidth}
 \begin{circuitikz}[scale=0.5 ]
 \draw  
  (7,2) node[xnor port,scale=0.5] (gx3) {3}
  (5,5.5) node[xor port,scale=0.5] (gx2) {4}
  (5,3.85) node[and port,scale=0.5] (ga2) {3}
  (3,3) node[xor port,scale=0.5] (gx1) {2}
  (3,1.5) node[nand port,scale=0.5] (ga1) {1}
 ;  
   \draw  
 (0,4.5) node[anchor=east] (a0){$p_0$} 
 (0,3.5) node[anchor=east] (a1){$p_1$} 
 (0,2.5) node[anchor=east] (a2){$p_2$} 
;    
\coordinate (a0s) at (0.5,5) ;
\coordinate  (a1s) at (0.5,1.5) ;
\coordinate (a2s) at (1.,2) ;

\draw 
  (a1) -| (gx1.in 1)
  (a2) -| (gx1.in 2)
  (a1) -| (a1s) (a1s) |- (ga1.in 2) 
  (a2) -| (a2s) (a2s) |- (ga1.in 1)  
  (a0) -| (a0s) (a0s) |- (gx2.in 1)
  (a0) -| (ga2.in 1)
 ;
 
 \draw 
 (gx1.out) -| (ga2.in 2)
 (gx1.out) |- (gx2.in 2)
 (ga1.out) -| (gx3.in 2)
 (ga2.out) -| (gx3.in 1)
 ;
 
 \draw 
  (gx2.out)   -- node[at end,right]{$s_0$} ++(right:22mm)
  (gx3.out)   -- node[at end,right]{$s_1$} ++(right:2mm)
  ;

 \end{circuitikz}
  \end{minipage}
 \hspace{0.4cm} 
 \begin{minipage}{0.2\textwidth}
\begin{center}
\begin{tabular}{ccc|cc} 
\toprule
$p_2$ & $p_1$ & $p_0$ & $s_1$ & $s_0$  \\
\midrule
0 & 0 & 0 & 0 & 0 \\
0 & 0 & 1 & 0 & 1 \\
0 & 1 & 0 & 0 & 1 \\
0 & 1 & 1 & 1 & 0 \\
1 & 0 & 0 & 0 & 1 \\
1 & 0 & 1 & 1 & 0 \\
1 & 1 & 0 & 1 & 0 \\
1 & 1 & 1 & 1 & 1 \\
\bottomrule
\end{tabular}
\end{center}
 \end{minipage}
\caption{ (Left) Classical gates computing 3-bit Hamming weight. Logical expressions:
$s_0=p_0 \oplus p_1 \oplus p_2 $, where $\oplus$ denotes {\it modulo 2} addition and 
$s_1= \overline{p_0(p_1 \oplus p_2) \oplus \overline{p_1 p_2}}$. The numbers inside classical gates map to equivalent quantum gates in \cref{fig:qbart-hammW}.
(Right) Truth table.
}
\label{fig:class-fpga2}
\end{figure}
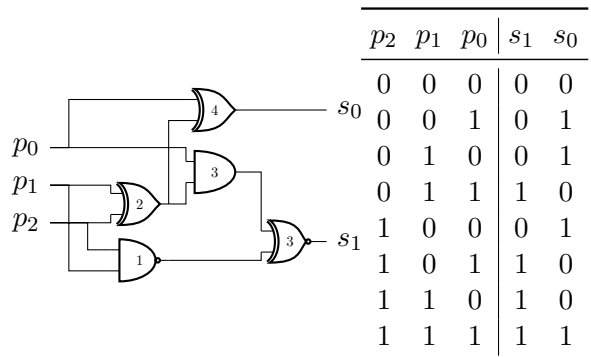

The Hamming weight of a bit-string is the number of 1s in the bit-string.
For completeness, we show the classical circuit computing the Hamming weight for a 3-bit input in \cref{fig:class-fpga2}.
It can be compared with the equivalent quantum circuit. To highlight again the analogy between classical and quantum gates, the numbers inside the classical gates in \cref{fig:class-fpga2} enumerate the equivalent quantum gates in \cref{fig:qbart-hammW}. It is easy to verify that the 4 quantum gates from \cref{fig:qbart-hammW} implement the 3-bit Hamming weight truth table shown here.

\section*{Acknowledgements}

All authors were supported by the Laboratory Directed Research and Development Program of Lawrence Berkeley National Laboratory and the Office of Advanced Scientific Computing Research under U.S. Department of Energy Contract No. DE-AC02-05CH11231.
 This research used resources of the National Energy Research Scientific Computing Center (NERSC), a U.S. Department of Energy Office of Science User Facility located at Lawrence Berkeley National Laboratory.

This research used resources of the Oak Ridge Leadership Computing Facility, which is a DOE Office of Science User Facility supported under Contract DE-AC05-00OR22725.

\bibliographystyle{abbrvurl}
\bibliography{references}

\newpage

\appendix

\section{Theoretical analysis of permuted $\UCRy$ gates}
\label{app:pUCRy}

In this section we provide a more detailed analysis of the construction of $\UCRy$ and the cyclically permuted $\UCRy$
circuits, illustrated in \cref{fig:pUCR}. The construction of regular $\UCRy$ circuits is well-understood~\cite{Mottonen2004, Amankwah2022QuantumImages}, but we include a brief discussion about these concepts for the sake of completeness.
The cyclically permuted $\UCRy$ and corresponding parallel $\UCRy$ circuits are a novel contribution to the best of our knowledge.

Throughout this text, two elementary properties of Pauli-$Y$ rotations will prove to be useful:
\begin{equation}
\begin{aligned}
\text{angle addition:} && R_y(\phi_0) R_y(\phi_1) &= R_y(\phi_0 + \phi_1),\\
\text{angle negation:} && X R_y(\phi) X &= R_y(-\phi).
\end{aligned}
\label{eq:rprops}
\end{equation}

\paragraph{Motivating example.}

We start with the case of a single address qubit as the simplest possible case to motivate the usage of the compact $\UCRy(\alpha)$ circuits. Using the decomposition for a singly-controlled $R_y$ gate into \cnot\ and single-qubit gates~\cite{NC2010},
\begin{equation}
\small
\quad
\begin{myqcircuit}
\lstick{a_0} & \ctrl{1} & \qw \\
\lstick{d}   & \gate{R_y(\alpha)} & \qw
\end{myqcircuit}
\quad=\qquad
\begin{myqcircuit}
\lstick{a_0} & \qw                  & \ctrl{1} & \qw                   & \ctrl{1} & \qw\\
\lstick{d}   & \gate{R_y(\alpha/2)} & \targ    & \gate{R_y(-\alpha/2)} & \targ    & \qw
\end{myqcircuit}\ ,
\end{equation}
which directly follows from the properties~\eqref{eq:rprops}.
We can decompose a $\UCRy$ circuit with a single address qubit as follows:
\begin{equation}
\small
\quad
\begin{myqcircuit}
\lstick{a_0} & \ctrlo{1} & \ctrl{1} & \qw \\
\lstick{d}   & \gate{R_y(\alpha_0)} & \gate{R_y(\alpha_1)} & \qw
\end{myqcircuit}
\quad = \qquad
\begin{myqcircuit}
\lstick{a_0} & \gate{X} & \qw                    & \ctrl{1} & \qw                     & \ctrl{1} & \gate{X} & \qw & \ctrl{1} & \qw                     & \ctrl{1} & \qw \\
\lstick{d}   & \qw & \gate{R_y(\alpha_0/2)} & \targ    & \gate{R_y(-\alpha_0/2)} & \targ    & \qw & \gate{R_y(\alpha_1/2)} & \targ    & \gate{R_y(-\alpha_1/2)} & \targ  & \qw
\end{myqcircuit}\ .
\end{equation}
This decomposition is clearly suboptimal as it is known that any two-qubit gate can be decomposed in a circuit with at
most 3 \cnot\ gates~\cite{Vidal04}. Using a compact $\UCRy$ circuit, we can implement the circuit using two \cnot\ gates only:
\begin{equation}
\small
\quad
\begin{myqcircuit}
\lstick{a_0} & \ctrlo{1} & \ctrl{1} & \qw \\
\lstick{d}   & \gate{R_y(\alpha_0)} & \gate{R_y(\alpha_1)} & \qw
\end{myqcircuit}
\quad = \qquad
\begin{myqcircuit}
\lstick{a_0} & \qw                  & \ctrl{1} & \qw                   & \ctrl{1} & \qw\\
\lstick{d}   & \gate{R_y(\theta_0)} & \targ    & \gate{R_y(\theta_1)} & \targ    & \qw
\end{myqcircuit}
\qquad,\qquad\qquad
\text{with}\qquad
\begin{aligned}
\alpha_0 &= \theta_0 + \theta_1 \\
\alpha_1 &= \theta_0 - \theta_1 \\
\end{aligned}.
\label{eq:circ_eq}
\end{equation}
Using~\eqref{eq:rprops}, we can verify that~\eqref{eq:circ_eq} holds by considering the action of the
circuit for the two possible basis states of the address qubit:
\begin{itemize}
\item If the address (control) qubit is in the $\ket{0}$ state, the circuit on the left applies a $R_y(\alpha_0)$
rotation to the second qubit. The circuit on the right applies $R_y(\theta_0)R_y(\theta_1) = R_y(\theta_0 + \theta_1)$
to the second qubit, which is equivalent if $\alpha_0 = \theta_0 + \theta_1$.
\item If the address (control) qubit is in the $\ket{1}$ state, the circuit on the left applies a $R_y(\alpha_0)$
rotation to the second qubit. The circuit on the right applies  $R_y(\theta_0)X R_y(\theta_1) X = R_y(\theta_0)R_y(-\theta_1) = R_y(\theta_0 - \theta_1)$ to the second qubit, which is equivalent if $\alpha_1 = \theta_0 - \theta_1$.
\end{itemize}
The relation between $\alpha_0, \alpha_1$ and $\theta_0, \theta_1$ is the Walsh-Hadamard transformation (\cref{eq:walsh} in the main document) of dimension $2$.

\paragraph{Cyclically permuted $\UCRy$ circuits.}
In the case of a single address qubit controlling $\UCRy$, $(n_a=1)$  , the decomposition~\eqref{eq:circ_eq} is \emph{unique}.
For $n_a > 1$, there are $n_a$ different realizations of the $\UCRy$ circuit that cyclically permute the index of the control qubit of the \cnot\ gates.
We illustrate this idea for $n_a = 2$. The first realization $\UCRy[2;0]$, with the \emph{permutation shift} $s = 0$ added as the 2nd superscript. For $s = 0$, we have the natural ordering of the address qubis  as $[0, 1]$:

\newcommand{\hl}[1]{%
  \colorbox{black!20}{$\displaystyle#1$}}

\begin{equation*}
\small{
\qquad
\begin{myqcircuit}
\lstick{a_0} & \qw &
         \ctrlo1 & \qw & \ctrlo1 & \qw &  \ctrl1 & \qw & \ctrl{1} & \qw \\
\lstick{a_1} & \qw &
         \ctrlo1 & \qw &  \ctrl1 & \qw & \ctrlo1 & \qw & \ctrl{1} & \qw \\
\lstick{d}   & \qw &
         \gate{R_y(\alpha_0)} & \qw & \gate{R_y(\alpha_1)} & \qw &
         \gate{R_y(\alpha_2)} & \qw & \gate{R_y(\alpha_3)} & \qw
\end{myqcircuit}
\quad = \qquad
\begin{myqcircuit}
\lstick{a_0} & \qw                  & \ctrl{2} & \qw                  & \qw      & \qw                  & \ctrl{2} & \qw                  & \qw      & \qw \\
\lstick{a_1} & \qw                  & \qw      & \qw                  & \ctrl{1} & \qw                  & \qw      & \qw                  & \ctrl{1} & \qw \\
\lstick{d} & \gate{R_y(\theta_0)} & \targ    & \gate{R_y(\theta_1)} & \targ    & \gate{R_y(\theta_2)} & \targ    & \gate{R_y(\theta_3)} & \targ    & \qw
\end{myqcircuit}
}
\end{equation*}
with
\begin{align*}
\alpha_0 &= \theta_0 + \theta_1 + \theta_0 + \theta_1, \\
\alpha_1 &= \theta_0 \hl{+} \theta_1 - \theta_0 \hl{-} \theta_1, \\
\alpha_2 &= \theta_0 \hl{-} \theta_1 - \theta_0 \hl{+} \theta_1, \\
\alpha_3 &= \theta_0 - \theta_1 + \theta_0 - \theta_1.
\end{align*}

The linear system relating $\alpha_i$'s to $\theta_j$'s can again be derived using~\eqref{eq:rprops}.
The second realization $\UCRy[2;1]$ cyclically permutes the position of the control of the $\cnot$ gates by 1, i.e., $s = 1$ which leads to the ordering of the address qubits as $[1, 0]$:

\begin{equation*}
\small{
\qquad
\begin{myqcircuit}
\lstick{a_0} & \qw &
         \ctrlo1 & \qw & \ctrlo1 & \qw &  \ctrl1 & \qw & \ctrl{1} & \qw \\
\lstick{a_1} & \qw &
         \ctrlo1 & \qw &  \ctrl1 & \qw & \ctrlo1 & \qw & \ctrl{1} & \qw \\
\lstick{d^{\prime}}   & \qw &
         \gate{R_y(\alpha_0)} & \qw & \gate{R_y(\alpha_1)} & \qw &
         \gate{R_y(\alpha_2)} & \qw & \gate{R_y(\alpha_3)} & \qw
\end{myqcircuit}
\quad = \qquad
\begin{myqcircuit}
\lstick{a_0}  & \qw                  & \qw      & \qw                  & \ctrl{2} & \qw                  & \qw      & \qw                  & \ctrl{2} & \qw \\
\lstick{a_1} & \qw                  & \ctrl{1} & \qw                  & \qw      & \qw                  & \ctrl{1} & \qw                  & \qw      & \qw \\
\lstick{d^{\prime}}   & \gate{R_y(\theta_0)} & \targ    & \gate{R_y(\theta_1)} & \targ    & \gate{R_y(\theta_2)} & \targ    & \gate{R_y(\theta_3)} & \targ    & \qw
\end{myqcircuit}}
\end{equation*}
with
\begin{align*}
\alpha_0 &= \theta_0 + \theta_1 + \theta_0 + \theta_1, \\
\alpha_1 &= \theta_0 \hl{-} \theta_1 - \theta_0 \hl{+} \theta_1, \\
\alpha_2 &= \theta_0 \hl{+} \theta_1 - \theta_0 \hl{-} \theta_1, \\
\alpha_3 &= \theta_0 - \theta_1 + \theta_0 - \theta_1.
\end{align*}

The only difference in the $s=1$ linear system that relates $\alpha_i$'s to $\theta_j$'s are the signs that are highlighted in gray.
This is essentially a permutation of the linear system for $s=1$ that can be computed efficiently by computing the position of the bit where
two consecutive Gray code $\texttt{g}_i$ and $\texttt{g}_{i+1}$ differ~\cite{Mottonen2004}, where $\texttt{g}_i$ is the reflected binary Gray code of the integer $i$.

This approach generalizes to any $n_a > 0$ address (control) qubits. There always exist $n_a$ different decompositions of the $\UCRy$ gate using the approach outlined above.  
The parameters $\theta_j$ can be computed from angles $\alpha_i$  using the \emph{Fast Walsh-Hadamard Transform} (FWHT)~\cite{Amankwah2022QuantumImages} followed by a permutation that depends on the shift $s$.
The classical complexity of this algorithm is $\bigO(n_a 2^{n_a})$.
As a $\pUCRy$ gate with $n_d$ data and $n_a$ address qubits is equivalent to $n_d$ permuted $\UCRy$ gates with $n_a$ address qubits, the cost of classical data preprocessing for \qcrank\ and \qbart\ thus is $\bigO(n_d n_a 2^{n_a})$.

\section{Further background on QPU benchmarks}
\label{app:QPU-benchmarks}

The benchmarks for \textit{Experiment \#5} as described in \cref{sec:QPUs-benchmarking} were executed on several cloud-accessible QPUs from Quantinuum, IonQ, and IBMQ. The basic device characteristics as reported at the time of the experiment are summarized in \cref{tab:ibmq-hw}. 
IBMQ devices do not support natively {\it all-to-all} connectivity, leading to significant overhead of CX-gates for the transpiled \qbart\ circuit. Consequently, IBMQ QPUs do not perform as good as a trapped ion QPU, even if single CX errors are comparable.

\begin{table}[hbtp]
\centering
\begin{tabular}{l|cccc} 
\toprule
& qubits & CX error & transpiled CX depth & \# sequences\\
\midrule
 Quantinuum H1-1 & 20 & 0.003 & 8 & 1-10\\
 IonQ Harmony & 11 & 0.040 & 8 & 10-15 \\
 IonQ Aria & 23  & 0.004 & 8 &  10 \\
 IBM guadalupe & 16 & 0.012 & 42 & 98 \\
IBM montreal & 27 & 0.020 & 39 & 98 \\
IBM jakarta  & 7 & 0.008 & 42 & 98 \\
\bottomrule
\end{tabular}
\caption{Basic characteristics of  quantum hardware  used for  \qbart\ \textit{experiment \#5} in \cref{sec:QPUs-benchmarking}. The listed average CX error is officially reported by the hardware providers at the time of circuit execution.
 }
 \label{tab:ibmq-hw}
\end{table}

\section{Further experiments using \qcrankabs\ and \qbartabs}
\label{app:experiments}

In addition to the five experiments on real hardware, as presented in \cref{sec:results}, we performed five other experiments using both \qcrank\ and \qbart\ using noise-free and noisy circuit simulators.
This additional data explores the robustness and versatility of proposed encodings and recovery techniques.
General information and setup for these additional experiments are summarized in \cref{tab:suppl_experiments}.
We use the same three metrics of fidelity as in \cref{sec:fidelity-metrics}: dynamic range ($D_r$), recovered value fidelity (RVF), and recovered sequence fidelity (RSF).
These metrics, along with our new {\it adaptive calibration} method, are described in \cref{sec:methods}.

\begin{table}[hbtp]
\centering
\begin{tabular}{lccccc} 
\toprule
Experiment & \#6 & \#7 & \#8 & \#9 & \#10 \\
		   & \qcrank\ & \qcrank\ & \qcrank\ & \qbart\ & \qbart\ \\   
\midrule
Data type & \multicolumn{4}{c}{Integer sequence} & Time series\\
Objective & \multicolumn{4}{c}{Test $D_r$, RVF, RSF} & ECG wave I/O \\
Simulator &  \multicolumn{4}{c}{ noisy$^\dag$ \texttt{Qiskit-Aer}  }  &  Quantinuum H1-1E$^\ddag$ \\
\cmidrule(lr){2-5}\cmidrule(lr){6-6}
addr. qubits ($n_a$) & 4 & 4 & 2 - 7 & 5 &  6 \\
data qubits ($n_d$) & 8 & 8 & 8 & 10 &  6 \\
\# addresses & 16 & 16 & 4 - 128 & 32 & 64\\
input (bits) & 384 & 384 & 96 - 3,000 & 320 & 384 \\
\bottomrule
\end{tabular}\\
\vspace{1mm}
{\footnotesize $^\dag$  \texttt{Qiskit} simulator with noise level tuned to approximate real QPUs.}\\
{\footnotesize $^\ddag$Quantinuum proprietary simulator H1-1E is tuned to emulate H1-1.}
\caption{Summary of simulated experiments using \qcrank\ and \qbart\ to provide further insights about the proposed encodings.}
\label{tab:suppl_experiments} 
\end{table}

We use \texttt{Qiskit-Aer} as the primary circuit simulator with four custom noise models, listed in \cref{tab:noise-model}, emulating the finite fidelities and coherence times of the QPUs that we used in the earlier experiments: (a) {\bf noise-free} ideal simulator for circuit validation and probing ideal performance, (b) {\bf minimal noise} levels to study the impact of introducing some noise over the ideal results, (c) {\bf H1-proxy} noise model that approximates the bit-flip and thermal noise present on the H1-1 Quantinuum trapped-ion QPU, (d) {\bf IBMQ-proxy} noise model based on the IBMQ transmon QPUs. 
The \texttt{Qiskit-Aer} simulator is set up to use {\it all-to-all} connectivity, which is not valid for IBMQ QPUs, and assumes no limit on the gates multiplicity per cycle, which is not valid for H1-1. 
Consequently, the CX-depth of the simulated circuits are shorter in comparison to the real hardware execution, as SWAPs are not required, and the fidelities retrieved from our simulations are better compared to the respective QPUs.

\begin{table}[hbtp]
\centering
\begin{tabular}{l|cccc} 
\toprule
{\bf Noise model} & ideal & minimal & H1-proxy & IBMQ-proxy\\
\midrule
{\bf Objective} & circuit verification & fidelity at low noise & \multicolumn{2}{c}{fidelity of  hardware }\\
\midrule
SPAM error$^\clubsuit$& 0 & $1\cdot 10^{-3}$ &  $3\cdot 10^{-3}$ &  $2.5 \cdot 10^{-2}$ \\
U3 error &  0 & $1\cdot 10^{-3}$ &  $5\cdot 10^{-5}$ &   $4.0\cdot 10^{-4}$\\
 CX error & 0 & $1\cdot 10^{-3}$ &  $3\cdot 10^{-3}$ &  $1.4 \cdot 10^{-2}$ \\
\midrule
duration T1/U3$^\spadesuit$ &  $\infty$ & $1\cdot 10^{5}$ &  $5000$ &   $2000$\\
duration T1/CX                &  $\infty$ & $1\cdot 10^{5}$ &  $170$ &   $200$\\
\midrule
RVF & 1.0 & 0.78 & 0.68 & 0.26\\
$D_r$ & 0.99 & 0.90 & 0.67 & 0.29\\
\bottomrule
\end{tabular}\\
\vspace{1mm}
$^\clubsuit$ bit flip probability error \\
$^\spadesuit$  thermal noise model relies  only the ratio of  coherence time to  gates duration.
 \caption{Noise model configurations used for simulated experiments listed in \cref{tab:suppl_experiments}.
 }
 \label{tab:noise-model} 
\end{table}

\subsection*{Experiment \#6: Adaptive calibration procedure for \qcrankabs}

The distortion correction function $g(\cdot)$, that was introduced in \cref{eq:the-rec-nisq}, is a heuristic 
calibration that can be
applied to the raw \qcrank\ data obtained from the QPU in order to improve the recovery fidelity compared to directly applying \cref{eq:frqi-rec} to the raw measurements.

The goal of this experiment is to study the performance of the adaptive calibration for the different noise models
listed in \cref{tab:noise-model}. 
We use a \qcrank\ circuit with 4 address- and 8 data-qubits such that the circuit has a depth of 32 CX-cycles and can
load $2^4 \times 8 = 128$ real values. 
We generate 98 random input data sequences of length 128 with values selected out of 8 different symbols, i.e, a bit-depth of 3.
The total capacity of this circuit configuration is 384 classical bits of information.
Each \qcrank\ circuit is measured for $3\cdot10^3$ shots for each noise model and every data set.

\begin{figure}[hbtp]
\centerline{\includegraphics[width=\textwidth]{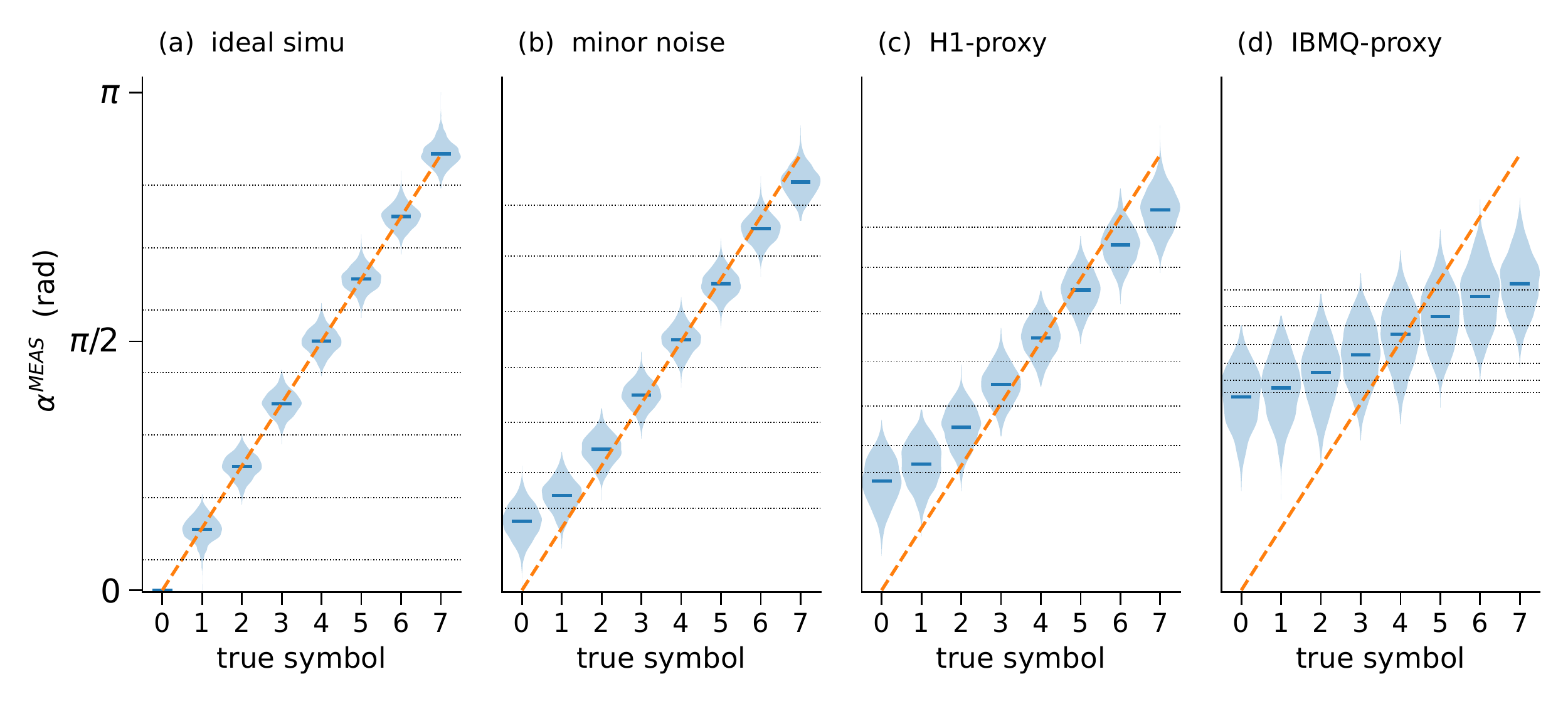}}
\caption{Reconstruction of $\alpha^{meas}$  (vertical axis) for \qcrank\  \textit{experiment \#6} with $3\cdot10^3$ shots and for four noise models listed in Table~\ref{tab:noise-model} are shown on panels (a)-(d). The true values of symbols (a.k.a. discrete inputs $\dpt$ ) are on the horizontal axis. The diagonal dashed red line marks the mathematically correct ideal reconstruction, \cref{eq:frqi-rec}, the same for all 4 panels.  The horizontal dotted lines mark adaptive calibration thresholds used for more accurate reconstruction following \cref{eq:the-rec-nisq}. The width of the violin plots denotes PDF  $\alpha^{meas}$.
}
\label{fig:sim-ang}
\end{figure}

\cref{fig:sim-ang} shows the distribution of the angles $\alpha^{meas}$, reconstructed using \cref{eq:frqi-rec} based on $3\cdot10^3$ shots obtained from the four different simulators, as a function of the input symbol. The reconstructed angles are visualized using \textit{violin plots} that show the distribution of the measured angles for each different input symbol. 
The average angle is indicated by the short blue bar.
We observe that for the ideal noise-free simulator, the measured averages line up exactly with the input angles that are shown as red dashed lines. Equivalently, the recovered angles will have no systematic bias if a large number of shots is used. Even for the ideal simulator there is a spread in the recovered angles caused by the finite sample size (shot noise). As the level of noise increases in panels (b)-(d), the position of the blue bars deviates more from the ideal red dashed line, and the spread on $\alpha^{meas}$ for an individual symbol increases.
The distortions correction function $g(.)$, defined in \cref{eq:g-alpha-meas}, is determined by the horizontal dotted lines
which indicate the heuristic intervals that map the measured angle to the most probable input symbol. The edges of these heuristic intervals are chosen as the average of two consecutive blue bars.

The lookup table for the distortion correction has to be precomputed through calibration, which requires additional shots,
but it can then be applied to remove the bias of new measurements on \qcrank\ circuits of data with similar characteristics.

The dynamic range ($D_r$), defined in \cref{eq:dynr}, is naturally visible in \cref{fig:sim-ang} as the vertical distance between the first and the last blue bars scaled to the full range of $\pi$.
The RVF, defined in \cref{sec:methods}, is the area of each violin that is contained between the respective threshold lines (dotted lines) assigned to a given symbol.
The final 2 rows in \cref{tab:noise-model} summarize $D_r$ and RVF obtained from the data shown in \cref{fig:sim-ang}.

\subsection*{Experiments \#7 and \#8: Fidelity of \qcrankabs\ as a function of shots and address qubits}

\begin{figure}[hbtp]
\centerline{\includegraphics[width=0.8\textwidth]{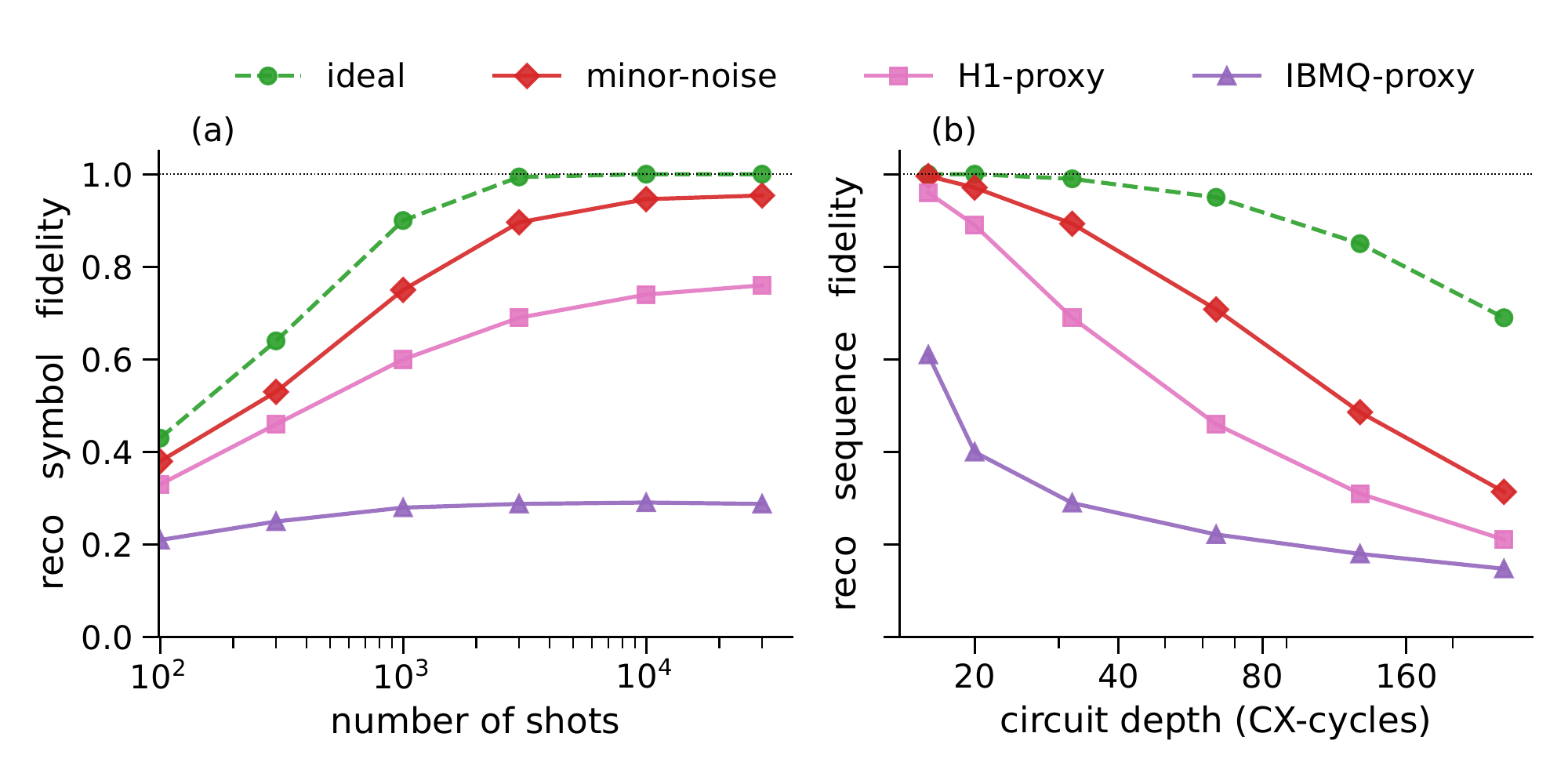}}
\vspace{-5pt}
\caption{Simulated \qcrank\ symbol fidelity for the 4 models of noise listed in \cref{tab:noise-model}. (a) \textit{experiment \#7}  shows fidelity dependence on the number of shots, while the problem size is constant. (b)  \textit{experiment \#8} shows fidelity as a function of  the size of the \qcrank\ addresses space, while the number of shots is kept constant.
}
\label{fig:sim-more}
\end{figure}

For {\it experiment \#7}, we use the same \qcrank\ setup as for the previous experiment 
but we vary the number of shots.
\cref{fig:sim-more}(a) shows the measured RVF as a function of the number of shots for the four different noise models.
The RVF improves with increasing number of shots, but the rate of improvement depends on the level of noise. 
Furthermore, with increasing noise, the RVF saturates at a lower fidelity and an RVF of 1 is only achieved for the noise-free simulator.
In other words, running more shots does not compensate the higher noise and sees a diminishing improvement in RVF. 
In \qcrank\ {\it experiment \#8}, we start again with the setup from {\it experiment \#6}, but keep the number of shots constant at $3\cdot10^3$.
Instead, we increase the number of address qubits, or equivalently the CX depth.
The measured RVF is shown in \cref{fig:sim-more}(b) as a function of the CX-cycles depth of the final circuit.
The RVF degrades with increasing address count due to the circuit being longer and there being fewer shots available per address.
Consequently, the measured probabilities are less accurate. This effect is more pronounced in the presence of noise.

\subsection*{Experiment \#9: Fidelity of \qbartabs\ as a function of shots} 
In this experiment, we perform a similar analysis for \qbart\ with majority voting suppressing the noise. The experimental setup is as follows: we use a \qbart\ circuit with 5 address qubits and 10 data qubits that has a information capacity of $2^5 \times 10 = 320$ classical bits. We use a random sequence of 320 bits and the same 4 noise models as before (see \cref{tab:noise-model}).

\begin{figure}[hbtp]
\centerline{\includegraphics[width=0.8\textwidth]{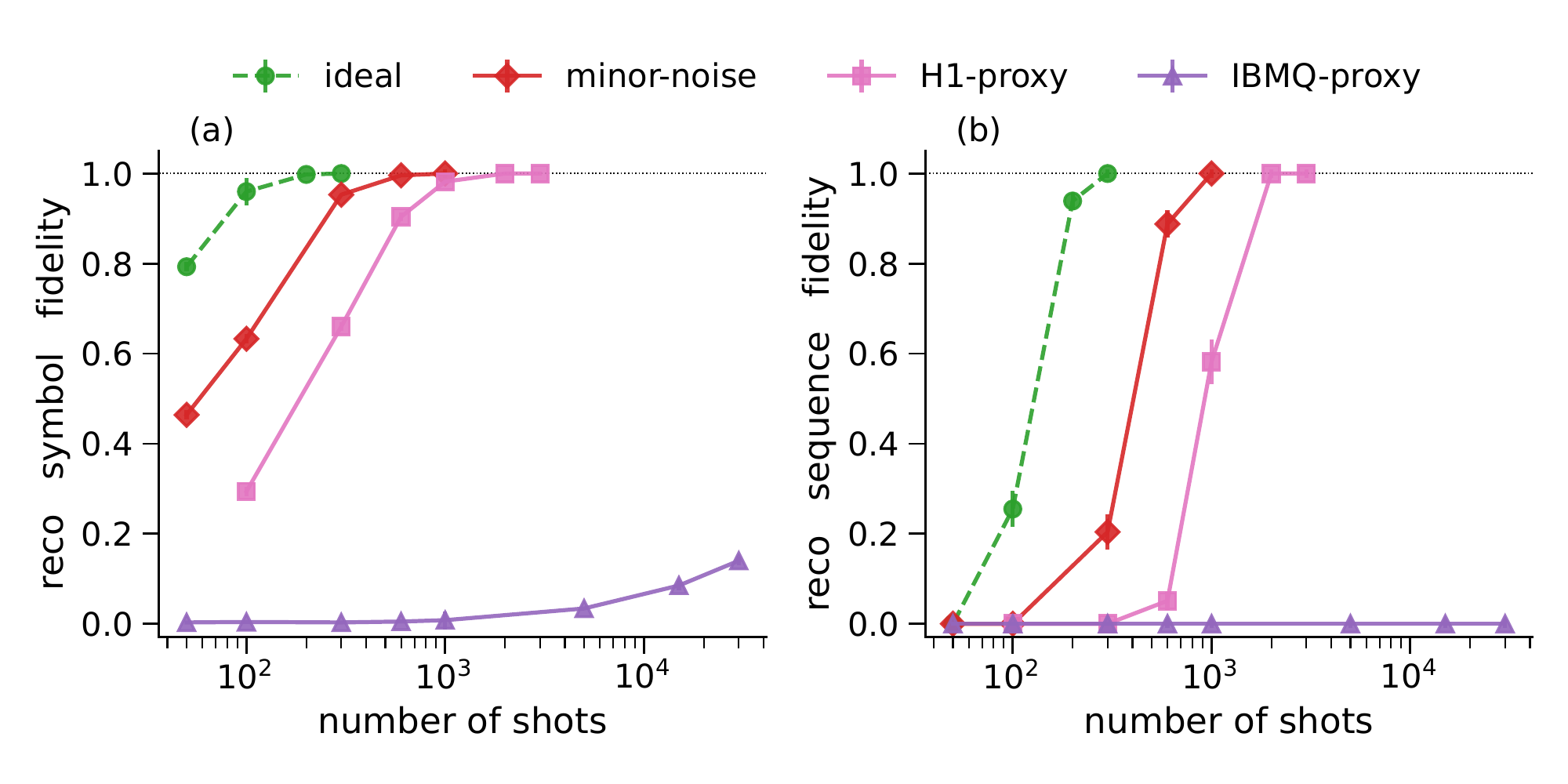}}
\vspace{-5pt}
\caption{Simulated \qbart\ reconstruction fidelity for \qbart\  \textit{experiment \#9} assuming various magnitued of the noise. (a) single value fidelity (b) whole sequence fidelity.
}
\label{fig:qbart-succ-shots}
\end{figure}

\cref{fig:qbart-succ-shots}(a) shows the RVF as a function of the number of shots.
For the noise-free, minimal noise, and H1-proxy models, the fidelity converges to $1.0$ for $\bigO(10^3)$ shots or fewer.
For the IBMQ-proxy noise model, the RVF remains low even if an order of magnitude more shots are used. With this noise model,
the bit strings are too corrupted for the majority voting technique to work.

\cref{fig:qbart-succ-shots}(b) shows the RSF, which measures that the \emph{full} sequence is retrieved correctly, again for the different noise models. This shows that a perfect recovery is possible using only a moderate number of shots as long as the noise-level is not too high. 

Comparing \cref{fig:qbart-succ-shots} to \cref{fig:sim-more} shows that \qbart\ requires significantly fewer shots to achieve a similar RVF. This is due to the sparser encoding and the data encoded in basis states, i.e. orthogonal states, compared to superpositions. Furthermore, as illustrated by the experiments in the main text, \qbart\ is more suitable for quantum data processing algorithms that act on the binary data representation, compared to the angle representation used in \qcrank.

\subsection*{Experiment \#10: ECG waveform time-series} 

\begin{figure}[hbtp]
\centerline{\includegraphics[width=.7\textwidth]{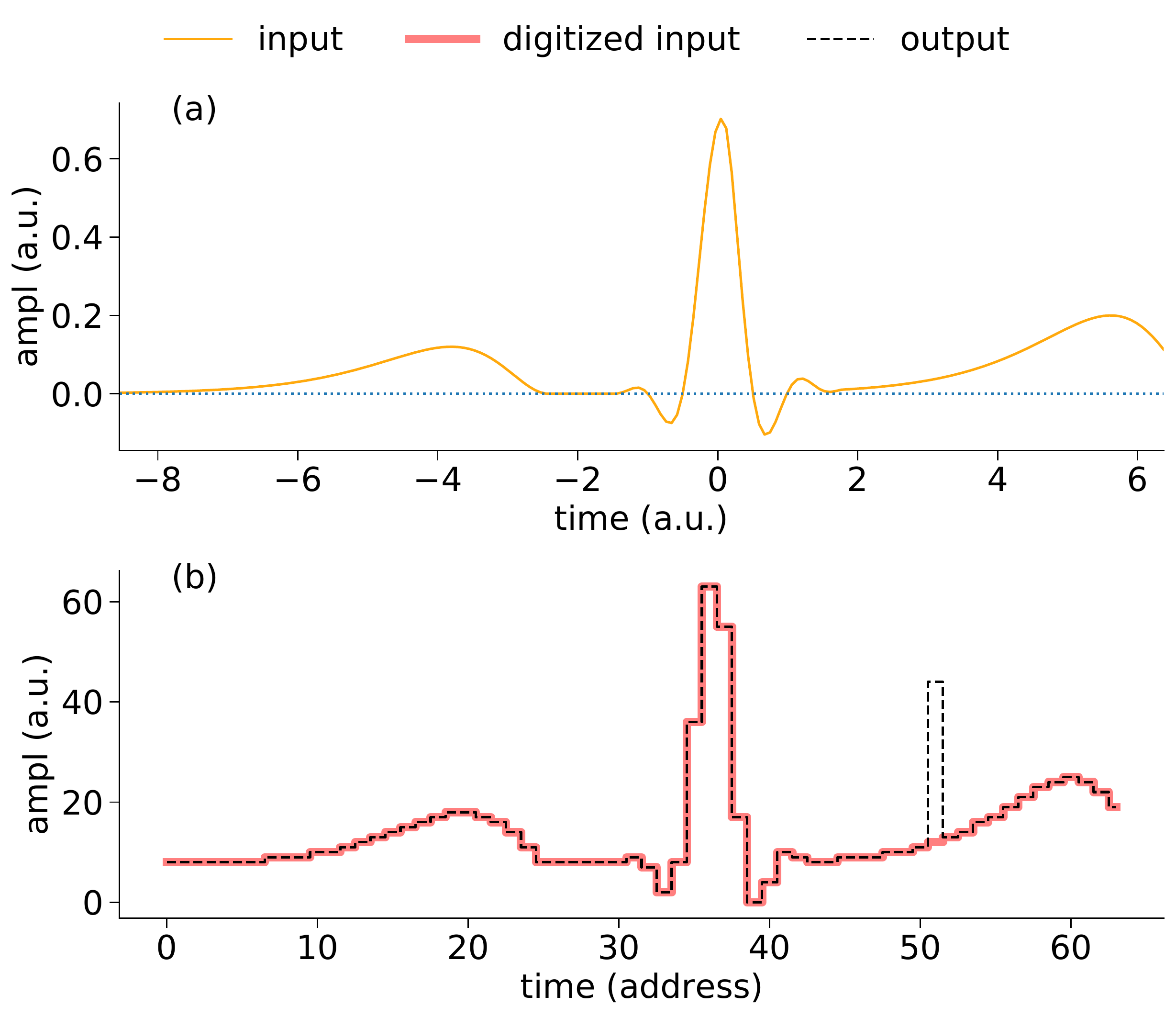}}
\caption{Encoding of an ECG signal using \qbart\ on the Quantinuum H1-1E emulator. The  12-qubit \qbart\ circuit is executed for $2\cdot10^3$ shots. (a) synthetic ECG signal. (b) digitized input is shown as a solid line and the reconstructed signal is presented as a dashed line.}
\label{fig:qbart-cardio}
\end{figure}

In this final experiment, we encode and recover a waveform
consisting of 64 values in 6-bit resolution using our \qbart\ encoding.
We generate a synthetic electrocardiogram (ECG) signal shown in \cref{fig:qbart-cardio}(a). 
This waveform is digitized into a sequence of 64 6-bit integers,
shown as solid red in \cref{fig:qbart-cardio}(b), and used as the \qbart\ input.
The dashed black line in the same figure shows the recovered ECG signal using majority voting.
The simulation is run on the Quantinuum H1-1E emulator and correctly recovers 63 out of the 64 input values
using 2000 shots.  
The \qbart\ circuit uses 6 address qubits and 6 data qubits and has the depth of the transpiled circuit is 64 CX gates.
Our other experiments suggest that the actual Quantinuum H1-1 QPU would deliver a similar performance using 150\% of the shots of the simulator.

\end{document}